\documentclass[aps,prb,twocolumn,superscriptaddress,showpacs]{revtex4-1}
\usepackage{graphicx,amsmath,color,placeins}
\let\hide\iffalse

\usepackage[caption=false]{subfig}

\begin{document}

\title{\textit{Ab initio} investigation on the experimental observation of metallic hydrogen}
\author{Xiaowei Zhang}
\affiliation{School of Physics and International Center for Quantum Materials, Peking University,
Beijing 100871, P. R. China}
\author{En-Ge Wang}
\email{egwang@pku.edu.cn}
\affiliation{School of Physics and International Center for Quantum Materials, Peking University,
Beijing 100871, P. R. China}
\affiliation{CAS Center for Excellence in Topological Quantum Computation, University of Chinese Academy of Sciences, Beijing 100190, P. R. China}
\affiliation{Collaborative Innovation Center of Quantum Matter, Peking University, Beijing 100871, P. R. China}
\author{Xin-Zheng Li}
\email{xzli@pku.edu.cn}
\affiliation{State Key Laboratory for Mesoscopic Physics and School of Physics, Peking University, Beijing 100871, P. R. China}
\affiliation{Collaborative Innovation Center of Quantum Matter, Peking University, Beijing 100871, P. R. China}
\date{\today}

\begin{abstract}
The optical spectra of hydrogen at $\sim$500 GPa were studied theoretically using a
combination of \emph{ab initio} methods.
Among the four most competitive structures, i.e. C2/c-24, Cmca-12, Cmca-4, and I41/amd,
only the atomic phase I41/amd can provide satisfactory interpretations of the recent experimental
observation, and the electron-phonon interactions (EPIs) play a crucial role.
Anharmonic effects (AHEs) due to lattice vibration are non-negligible but not sufficient to
account for the experimentally observed temperature dependence of the reflectance.
The drop of the reflectance at 2 eV is not caused by diamond's band gap reducing or interband plasmon,
but very likely by defects absorptions in diamond.
These results provide theoretical support for the recent experimental realization of metallic hydrogen.
The strong EPIs and the non-negligible AHEs also emphasize the necessity for quantum treatments
of both the electrons and the nuclei in future studies.
\end{abstract}
\maketitle
\indent
\section{Introduction}
Ever since the prediction of Wigner and Huntington in 1935 that metallic hydrogen (MH) would
form at high pressures,~\cite{wigner1935possibility} the search for MH has ranged among the biggest
challenges in condensed matter physics and high-pressure physics.~\cite{ashcroft1968metallic,loubeyre1996x,narayana1998solid,goncharov2001spectroscopic,
bonev2004quantum,Pickard2007,mcmahon2011ground,zha2012synchrotron,chen2013quantum,azadi2014dissociation,dias2017observation,McMinis2015}
In 1968, considering the fact that hydrogen (H) is the lightest among all elements and
the electron-phonon coupling could be strong in MH, Ashcroft proposed that this MH is
a candidate of high-temperature ($T$) superconductor.~\cite{ashcroft1968metallic}
In recent years, with the advent of several \emph{ab initio} methods, the superconductor
behavior of MH has been thoroughly studied in atomistic detail.~\cite{cudazzo2008ab,borinaga2016anharmonic,mcmahon2011high}
Other interesting phenomena, e.g. the low-$T$ metallic liquid phase, the superfluid phase due
to nuclear quantum effects (NQEs), and the potential rocket fuel properties, were also reported.~\cite{bonev2004quantum,babaev2004superconductor,chen2013quantum,silvera2010metallic}
Despite all these intriguing theoretical proposals, the experimental realization of the MH which
underlies the existence of all these exciting and non-trivial phenomena, remains ambiguous.
As such, the experimental verification of MH is nowadays commonly viewed as the \emph{Holly Grail} in
high-pressure physics.~\cite{Ceperley2012}\\
\indent
Due to the small scattering cross-section of H to X-ray and electron beams, except for some
extremely challenging experiments,~\cite{mao1988synchrotron,loubeyre1996x,akahama2010evidence}
most experimental characterizations of the crystal structures of H in the $\sim$100 GPa and higher pressure
range resort to indirect lattice vibration measurements such as
the infrared (IR) and the Raman spectroscopy.~\cite{lorenzana1989evidence,zha2012synchrotron,hanfland1993novel,lorenzana1990orientational,hanfland1992synchrotron,goncharov2001spectroscopic,
hemley1990low,goncharov1996raman,hemley1997vibron,gregoryanz2003raman,zha2013high,dalladay2016evidence,howie2015raman,zha2014raman}
Concerning the electronic structures, direct measurement of the conductivity and band structure is
also difficult.~\cite{drozdov2015conventional,eremets2011conductive,eremets2017molecular}
As such, the insulator-to-metallic phase transition was often characterized by visual optical
observations.~\cite{mao1990infrared,eggert1991absorption,hemley1991high}
Based on the observation that H turns nearly opaque at $\sim$250 GPa, Mao and Hemley claimed the
first low-$T$ MH using experimental evidence of metallization by the band
overlap in 1989.~\cite{mao1989optical}
A consensus, however, was not reached on this observation and a series of continuous efforts
were reported by different experimental groups.~\cite{howie2012mixed,zha2012synchrotron,goncharov2001spectroscopic,loubeyre2002optical,evans1998index,narayana1998solid}
During this time, the progress of \textit{ab initio} methods, the crystal structure searching methods in
particular,~\cite{oganov2006crystal,pickard2006high,wang2010crystal} has enabled a detailed atomistic theoretical understanding of the insulator-to-metal phase
transition.~\cite{Pickard2007,pickard2012density,liu2012quasi,mcmahon2011ground}
Different calculations indicate that H may become metallic during the pressure
range 350 to 500 GPa.~\cite{mcmahon2011ground,McMinis2015,azadi2014dissociation}
Most recently, Dias and Silvera (DS) announced that they observed atomic metallic hydrogen in
2017.~\cite{dias2017observation}
At 495 GPa and low $T$s, the reflectance of this MH is as high as 0.91.~\cite{dias2017observation}
Debates concerning the pressure calibration and the reflectance measurement soon
arise.~\cite{eremets2017comments,loubeyre1702comment,silvera2017response,goncharov2017comment,liu2017comment,silvera2017science,geng2017public}
Reproduction of the DS's experiment and extensions beyond that are clearly necessary for a final
confirmation of the MH from the experimental perspective.
From the theoretical side, direct $\emph{ab initio}$ simulations of the reflectance will also
help to understand the changes of the atomic structures and the electronic structures happening
at this pressure range.~\cite{borinaga2018strong}\\
\indent
In this paper, we investigate the optical spectra of MH at this pressure range by comparing
directly the reflectance with experiments.
Four candidate structures, i.e. C2/c-24, Cmca-12, Cmca-4, and I41/amd, were chosen.
The structures are labelled by their short Hermann-Mauguin space-group symbols, and the numbers
are additional information meaning the number of atoms in the primitive unit cell to avoid ambiguity.
These four solid phases were considered as the most competitive ones at 300 to 500 GPa in
terms of static enthalpy, and when the zero-point energy (ZPE) corrections were included.
\emph{Ab inito} density-functional theory (DFT) in combination with semiclassical Frank-Condon (SCFC)
principles were used to describe the optical spectra with the influence of electron-phonon
interactions (EPIs) included.
Among these four structures, only the reflectance of the atomic metallic hydrogen I41/amd can give a
satisfactory explanation of the experimental observations.
The influence of nuclear anharmonic effects (AHEs) on the spectra is non-negligible.
But it is not sufficient to account for the $T$-dependence of the experimental observed
reflectance.
Therefore, the $T$-dependence of the original experimental data is very likely to be extrinsic to H.
Our calculations also show that the drop of the reflectance at 2 eV is not caused by the diamond's
band gap reducing or the interband plasmon.
Rather, correcting the calculated reflectance using experimental absorption data of diamond's
defects can reproduce the reflectance drop above 2 eV.
Combining these results, we provide a theoretical support for the recent DS's experimental realization
of MH.
Analysis of the EPIs indicates that in future studies static treatment of the nuclei is far from
being enough in describing the optical and electronic structures of this material.
A fully quantum treatment of both the electrons and nuclei with AHEs taken into account, therefore,
will often be needed.\\
\indent
The paper is organized as follows.
The methods and computational details were explained in Sec. II.
The results and discussions were presented in Sec. III.
In particular, we focus on the influence of EPIs on the optical spectra accessible to such
ultrahigh-pressure experiments, and compare the reflectance with the available ones.
The conclusion was given in Sec. IV.
\section{Methods and computational details}
\subsection{Methods}
The linear optical properties of crystals are characterized by the long-wavelength macroscopic
dielectric function:
\begin{equation}
\varepsilon^{\text{M}}(\hat{\mathbf{q}},\omega)=\sum_{\alpha\beta}\hat{\mathbf{q}}_{\alpha}\varepsilon^{\text{M}}_{\alpha\beta}(\omega)\hat{\mathbf{q}}_{\beta},
\end{equation}
where $\hat{\mathbf{q}}=\mathbf{q}/|\mathbf{q}|$ is the unit wave vector of the incident light.
$\varepsilon^{\text{M}}$ is complex and its real part and imaginary part are related to the refractive index and extinction coefficient through:
\begin{eqnarray}
n_{\alpha\alpha}(\omega)=\sqrt{\frac{|\varepsilon^{\text{M}}_{\alpha\alpha}(\omega)|+\text{Re}\, \varepsilon^{\text{M}}_{\alpha\alpha}(\omega)}{2}},
\label{refractive}
\\
\kappa_{\alpha\alpha}(\omega)=\sqrt{\frac{|\varepsilon^{\text{M}}_{\alpha\alpha}(\omega)|-\text{Re}\, \varepsilon^{\text{M}}_{\alpha\alpha}(\omega)}{2}}.
\label{extinction}
\end{eqnarray}
The reflectance at normal incidence is then calculated by:
\begin{equation}
R_{\alpha\alpha}(\omega)=\frac{(n_{\alpha\alpha}-1)^2+\kappa_{\alpha\alpha}^2}{(n_{\alpha\alpha}+1)^2+\kappa_{\alpha\alpha}^2}.
\label{reflect}
\end{equation}
It should be noted that the $n$, $\kappa$ and $R$ obtained are only defined for the
diagonal dielectric tensor.~\cite{ambrosch2006linear}\\
\subsubsection{Static dielectric function}
\indent
The macroscopic dielectric tensor is linked to the microscopic inverse dielectric matrix by:
\begin{equation}
\varepsilon^{\text{M}}(\hat{\mathbf{q}},\omega)=\frac{1}{\lim\limits_{\mathbf{q} \to 0}[\varepsilon^{-1}(\mathbf{q},\omega)]_{\mathbf{G}=0,\mathbf{G}^{\prime}=0}},
\end{equation}
where $\mathbf{G}$ and $\mathbf{G}^{\prime}$ are reciprocal lattice vectors.
Usually, the random-phase approximation (RPA) is adopted in describing the dielectric
matrix,~\cite{gajdovs2006linear} using:
\begin{equation}
\varepsilon_{\mathbf{G},\mathbf{G}^{\prime}}(\mathbf{q},\omega)=\delta_{\mathbf{G},\mathbf{G}^{\prime}}-\frac{4\pi e^2}{|\mathbf{G}+\mathbf{q}||\mathbf{G}^{\prime}+\mathbf{q}|}\chi_{\mathbf{G},
\mathbf{G}^{\prime}}^{0}(\mathbf{q},\omega).
\end{equation}
$\chi^{0}(\mathbf{q},\omega)$ is the so-called independent-particle irreducible polarizability,
because under RPA the system's response to the total field (induced and incident field) resembles the case of non-interacting systems.~\cite{aryasetiawan1998gw}\\
\indent
If one neglects the local field effects,~\cite{louie1975} i.e. contributions from the
off-diagonal matrix elements of $\varepsilon_{\mathbf{G},\mathbf{G}^{\prime}}(\mathbf{q},\omega)$
to its inverse matrix, one has:
 \begin{equation}
 \varepsilon^{\text{M}}(\hat{\mathbf{q}},\omega)={\lim\limits_{\mathbf{q} \to 0}\varepsilon_{0,0}(\mathbf{q},\omega)}.
 \end{equation}
This approximation is the so-called ``neglecting local filed effects'' and also referred to as
independent particle approximation (IPA).
In so doing, the imaginary part of macroscopic dielectric function can be obtained using the
Kohn-Sham orbitals and eigenvalues,~\cite{harl2008linear} by:
\begin{equation}
\begin{split}
\text{Im}\,\varepsilon_{\alpha\beta}(\omega)=&\frac{4\pi^2e^2}{V}\lim\limits_{\mathbf{q}\to0}\frac{1}{q^2}\sum_{nm\mathbf{k}} 2f_{n\mathbf{k}}<u_{m\mathbf{k}+\mathbf{e}_{\alpha}\mathbf{q}}|u_{n\mathbf{k}}>\times\\
&<u_{n\mathbf{k}}|u_{m\mathbf{k}+\mathbf{e}_{\beta}\mathbf{q}}>[\delta(\epsilon_{m\mathbf{k}}-\epsilon_{n\mathbf{k}}-\hbar\omega)-\\
&\delta(\epsilon_{m\mathbf{k}}-\epsilon_{n\mathbf{k}}+\hbar\omega)],
\end{split}
\label{imagepsilon}
\end{equation}
where $u_{n\mathbf{k}}$ and $u_{m\mathbf{k}}$ are the periodic parts of the Bloch wave functions
for the initial and final states, $\epsilon_{n\mathbf{k}}$ and $\epsilon_{m\mathbf{k}}$ are the
eigenvalues, and $V$ is the volume of the unit cell.
$f_{n\mathbf{k}}$ is the Fermi occupation number and 2 comes from the sum over spin.
The real part of the dielectric function can be obtained by the Kramers-Kr\"onig transformation.
The $n\neq m$ terms in Eq.~\ref{imagepsilon} contribute to the interband transitions while the
$n=m$ term contributes to the intraband transitions.
The latter exists only in metals and vanish at non-zero frequency when the electron-electron
and electron-phonon interactions are neglected.
With EPIs, it extends to non-zero frequency which can be described empirically using a relaxation time
$\tau$ or within an \textit{ab initio} framework as we will discuss in Sec. II A.2.
As such, the optical properties of metals are largely affected by the real part of the intraband
dielectric functions,~\cite{harl2008linear} with the form:
\begin{equation}
\text{Re}\,\varepsilon_{\alpha\beta,\text{intra}}(\omega)=-\frac{\omega^{2}_{\text{p},\alpha\beta}}{\omega^2},
\label{realintra}
\end{equation}
where $\omega_{\text{p},\alpha\beta}$ is the plasma frequency and it often needs large number of
$\mathbf{k}$-points to converge.
In so doing, the macroscopic dielectric functions in metals can be clearly separated into two
terms, as:
\begin{equation}
\varepsilon_{\alpha\beta}(\omega)=\varepsilon_{\alpha\beta,\text{intra}}(\omega)
+\varepsilon_{\alpha\beta,\text{inter}}(\omega).
\end{equation}
\subsubsection{William-Lax (WL) method}
\indent
In many theoretical simulations, the dielectric functions in Eq.~\ref{imagepsilon} are calculated with
static nuclei clamped at the equilibrium positions.
With this treatment, the EPIs are completely neglected.
To include the effects of EPIs on the optical spectra, one can start from the Fermi's Golden rule,
which states that the optical transition rate from an initial quantum state $\Psi_{i}$ to a final quantum
state $\Psi_{f}$ can be calculated by:
\begin{equation}
W_{fi}(\omega)=\frac{2\pi}{\hbar}|<\Psi_{f}|\hat{M}|\Psi_{i}>|^2\delta(E_{f}-E_{i}-\hbar\omega).
\label{tranrate}
\end{equation}
Here $\hat{M}$ is the perturbation Hamiltonian, $i$ and $f$ refer to the quantum numbers of the initial
and final states of the electron-nuclei coupled quantum system.
Using the concepts of Born-Oppenheimer adiabatic (BOA) approximation and Born-Oppenheimer potential
energy surface (BO-PES), the total wave function can be represented as a product of the electronic part
and the nuclear part, as:
\begin{equation}
\Psi_{kn}(\mathbf{r},\mathbf{R})\approx\phi_{k}(\mathbf{r},\{\mathbf{R}\})\chi_{kn}(\mathbf{R}).
\label{wvfunction}
\end{equation}
$\mathbf{r}$($\mathbf{R}$) means the electronic (nuclear) coordinates.
$\phi_{k}$ is the $k^{\text{th}}$ electronic eigenstate, determined parametrically by the nuclear
configuration $\{\mathbf{R}\}$.
$\chi_{kn}(\mathbf{R})$ means the $n^{\text{th}}$ nuclear eigenstate on the $k^{\text{th}}$ electronic
BO-PES.\\
\indent
Substituting the $\Psi_{i}$ and $\Psi_f$ in Eq.~\ref{tranrate} by the form of $\Psi_{kn}$
in Eq.~\ref{wvfunction}, we obtain the quantum-mechanical transition rate:
\begin{equation}
\begin{split}
W_{ba}(\omega)=\frac{2\pi}{\hbar}\frac{1}{Z_{a}}\sum_{mn}e^{-\beta E_{am}}&|<\chi_{bn}|M_{ba}|\chi_{am}>|^2\times\\
&\delta(E_{bn}-E_{am}-\hbar\omega).
\end{split}
\label{tranrate2}
\end{equation}
$a$ and $b$ label the electronic states.
$n$ and $m$ label the nuclear (vibronic) states.
Starting from the electronic state $a$, $Z_{a}=\sum_{m}e^{-\beta E_{am}}$ is the partition
function to address nuclear motion on this electronic BO-PES.
$M_{ba}=<\phi_{b}|\hat{M}|\phi_{a}>$ is the dipole matrix element between the electronic states $a$
and $b$.\\
\indent
The calculation of Eq.~\ref{tranrate2} is challenging for two reasons.
On one hand, e.g. in crystals, the vibrational modes are enormous and thus the computational cost
of integrals between two vibrational states are very large.
On the other hand, the requirement of knowledge about the vibrational states on the excited
electronic states is also a huge challenge.
To simplify the problem, Lax proposed a semiclassical form (SCFC principle) in his seminal paper.~\cite{lax1952franck}
The main idea is to replace the difference between the two discrete total energy $E_{bn}$ and $E_{am}$ by the BOA classical potential energy which depends parametrically on the nuclear configuration,
i.e. $E_{bn}-E_{am}\approx U_{b}(\mathbf{R})-U_{a}(\mathbf{R})$.
After this, the vibrational quantum numbers of the final electronic state disappear.
And the transition rate simplifies into:
\begin{equation}
W_{ba}^{\text{SC}}(\omega)=\frac{1}{Z_{a}}\text{Tr}\hat{\rho}_{a}P_{ba}
\end{equation}
with
\begin{equation}
P_{ba}(\omega;\mathbf{R})=\frac{2\pi}{\hbar}|M_{ba}(\mathbf{R})|^2\delta (U_{b}(\mathbf{R})-U_{a}(\mathbf{R})-\hbar\omega).
\end{equation}
$\rho_{a}=e^{-\beta\hat{H}_{a}}$ is the density operator.
Note that the imaginary part of dielectric function is obtained by summing over all final electronic states,~\cite{patrick2014unified} i.e. Im~$\varepsilon(\omega;\mathbf{R})\propto\sum_{b}(1/\omega) P_{0b}(\omega;\mathbf{R})$ and Re~$\varepsilon(\omega;\mathbf{R})$ is related to
Im~$\varepsilon(\omega;\mathbf{R})$ by the Kramers-Kr\"onig transformation. So when including EPIs, the dielectric function is modified as
\begin{equation}
\varepsilon(\omega,T)=\frac{1}{Z_{0}}\text{Tr}\hat{\rho}_{0}\varepsilon(\omega;\mathbf{R}),
\label{finiteTepsilon}
\end{equation}
where 0 means the electronic ground state.
The key to calculate Eq.~\ref{finiteTepsilon} lies in the sample of the nuclear configurational space
which can be treated numerical, e.g. through the path-integral Monte-Carlo (PIMC) or the
path-integral molecular dynamics (PIMD) methods.\\
\indent
The underlying principle of PIMC and PIMD is that the finite-$T$ dielectric function in
Eq.~\ref{finiteTepsilon} can be rewritten as:
~\cite{della2004quantum,tuckerman2010statistical,enge2018computer}
\begin{equation}
\begin{split}
\varepsilon(\omega,T)=\lim\limits_{P\to\infty}&\frac{1}{Z_{P}}\prod_{j=1}^{N}(\frac{m_{j}P}{2\beta\pi\hbar^{2}})^{\frac{P}{2}}\int_{V}\int_{V}\cdots\int_{V} \\
&e^{-\beta \sum\limits_{i=1}^{P}[\sum\limits_{j=1}^{N}\frac{m_{j}}{2}\omega_{P}^2(\mathbf{x}_{i}^{j}-\mathbf{x}_{i-1}^{j})^2+\frac{1}{P}U_{0}(\mathbf{x}_{i}^1,\cdots,\mathbf{x}_{i}^{N})]}\\
&\times\frac{1}{P}\sum\limits_{i=1}^{P}\varepsilon(\omega;\{\mathbf{x}_{i}\})d\mathbf{x}_{1}d\mathbf{x}_{2}\cdots d\mathbf{x}_{P}.
\end{split}
\label{piepsilon}
\end{equation}
Here P is the number of imaginary time slices.
$\mathbf{x}_{i}$ means the nuclear configuration of the $i^{\text{th}}$ slice of the $N$-atom system.
$m_{j}$ is the mass of the $j^{\text{th}}$ nucleus.
$Z_{P}$ is the partition function of the classical isomorphic polymer, which equals:
\begin{equation}
\begin{split}
Z_P=&\prod_{j=1}^{N}(\frac{m_{j}P}{2\beta\pi\hbar^{2}})^{\frac{P}{2}}\int_{V}\int_{V}\cdots\int_{V} \\
&e^{-\beta \sum\limits_{i=1}^{P}[\sum\limits_{j=1}^{N}\frac{m_{j}}{2}\omega_{P}^2(\mathbf{x}_{i}^{j}-\mathbf{x}_{i-1}^{j})^2+\frac{1}{P}U_{0}(\mathbf{x}_{i}^1,\cdots,\mathbf{x}_{i}^{N})]}d\mathbf{x}_{1}d\mathbf{x}_{2}\cdots d\mathbf{x}_{P}.
\end{split}
\label{partitionfunct}
\end{equation}
$\omega_{P}=\sqrt{P}/(\beta\hbar)$ determines the strength of spring interactions between the
neighboring slices.
In our simulations, we have employed the PIMD method for the sampling of the statistic NQEs along the
imaginary time axis.
We label this method for the calculation of the spectra as WL-PIMD in the later discussions.\\
\indent
This WL-PIMD method rigorously accounts for the NQEs including the AHEs.
But we note that in practical simulations, the finite-$T$ dielectric function in
Eq.~\ref{finiteTepsilon} can also be calculated within the harmonic approximation (HA), in
which the computational cost is substantially reduced due to the analytical nature of the
harmonic phonon wave functions.
This method is labelled as the WL-HA method in our discussions.
In Ref.~\onlinecite{zacharias2016one}, Zacharias and Giustino have shown that
Eq.~\ref{finiteTepsilon} can be rewritten as a more compact form:
\begin{equation}
\varepsilon(\omega,T)=\prod_{\nu}\int dx_{\nu}\frac{e^{-x_{\nu}^{2}/2\sigma_{\nu,T}^{2}}}{\sqrt{2\pi}\sigma_{\nu,T}}\varepsilon(\omega;\mathbf{R}),
\label{wlmethodharm}
\end{equation}
with
\begin{eqnarray}
\sigma_{\nu,T}^{2}=(2n_{\nu,T}+1)l_{\nu}^2\\
l_{\nu}=\sqrt{\frac{\hbar}{2m_{0}\omega_{\nu}}}.
\end{eqnarray}
Here, $x_{\nu}$ denotes norm mode coordinates, $\omega_{\nu}$ is the phonon frequency, $m_0$ is
the reference mass and they have chosen as that of a proton, and $n_{v,T}$ is Bose-Einstein
occupation number.\\
\indent
Eq.~\ref{wlmethodharm} can be evaluated efficiently using Monte Carlo integration
techniques.~\cite{zacharias2015stochastic}
Specifically, a set of norm mode coordinates can be generated randomly, which correspond to a set of
atomic structures.
When these randomly chosen atomic structures present a complete sampling of the finite-$T$
atomic configurations, Eq.~\ref{wlmethodharm} provides a rather accurate description of the optical
spectra with EPIs taken into account within HA.
More recently, Zacharias and Giustino further demonstrated that Eq.~\ref{wlmethodharm} can be evaluated
rather accurately using only 1-2 configurations of the nuclei.~\cite{zacharias2016one}
The main idea is to take the value of norm mode coordinates to be $\pm\sigma_{\nu,T}$, i.e.
\begin{equation}
\Delta\tau_{\kappa\alpha}=\sqrt{\frac{m_{0}}{m_{\kappa}}}\sum\limits_{\nu}(-1)^{\nu-1}e_{\kappa\alpha,\nu}\sigma_{\nu,T},
\label{distortion}
\end{equation}
where $m_{\kappa}$ is the mass of the $\kappa^{\text{th}}$ atom, and $e_{\kappa\alpha,\nu}$ is the phonon eigenmode.
\subsection{Computational details}
\indent
The geometry optimizations, the \textit{ab initio} PIMD
simulations,~\cite{marx1994ab,marx1996ab,tuckerman1996efficient}
the WL-HA simulations of the optical spectra,~\cite{zacharias2015stochastic,zacharias2016one}
and the WL-PIMD simulations of the optical spectra
were performed using the Vienna \emph{Ab initio} Simulation Package (VASP).~\cite{kresse1996efficiency,kresse1996efficient}
Projector augmented-wave (PAW) potential and Perdew-Burke-Ernzerhof (PBE) exchange-correlation
functional were chosen.~\cite{blochl1994projector,kresse1999ultrasoft,perdew1996generalized}
In the \textit{ab initio} PIMD simulations, the canonical ensemble (NVT) was employed and the simulations
were performed at 83 K and 50 K with 64 beads, together with a supercell containing 72 atoms.
The timestep was 0.5 fs and the simulation is 5 ps long.
10 uncorrelated snapshots from the trajectory are chosen to calculate the optical spectra.
With 64 beads per snapshot, 64$\times$10 optical spectra calculations are needed in order
to obtain the WL-PIMD results.
Analysis on convergence shows that these settings are enough.
For the WL-HA calculations, phonon dispersions were obtained using the finite-displacement method
with Phonopy package,~\cite{togo2015first} and the supercell for I41/amd, Cmca-4, Cmca-12 and C2/c-24 includes 108, 216, 96 and 432 atoms, respectively.
Then two distorted atomic structures were produced using Eq.~\ref{distortion} and
supercells containing 864, 512, 768 and 648 atoms were used respectively to converge the WL-HA results.
The calculations of optical spectra were carried out within DFT-IPA.
For geometry optimized structure with static nuclei, the reflectance of
I41/amd were calculated using a $110\times110\times110$ $\mathbf{k}$-mesh for the
Brillouin-Zone (BZ) sampling.
For WL-HA, 40, 240, 20 and 80 random $\mathbf{k}$-points were used for I41/amd, Cmca-4, Cmca-12 and C2/c-24, respectively.
4000 random $\mathbf{k}$-points were used to calculate the reflectance of I41/amd for 72-atom
supercell under HA.
For WL-PIMD, 1000 random $\mathbf{k}$-points were used to converge the reflectance.
The effective band structures were obtained using a 256-atom supercell.
Diamond's Raman spectra were calculated with density-functional perturbation theory (DFPT) using Quantum Espresso,~\cite{baroni2001phonons,giannozzi2009quantum}
with local-density approximation (LDA) exchange-correlational functional and norm-conserving
pseudopotential used.
Diamond's band gap was obtained using the $G_{0}W_{0}$ method upon LDA in
VASP.~\cite{hedin1965, hybertsen1986, shishkin2006, shishkin2007}
\section{Results and Discussions}
\subsection{Electron-Phonon Interactions}
\begin{figure*}[ht!]
	\subfloat{\includegraphics[width=0.39\textwidth]{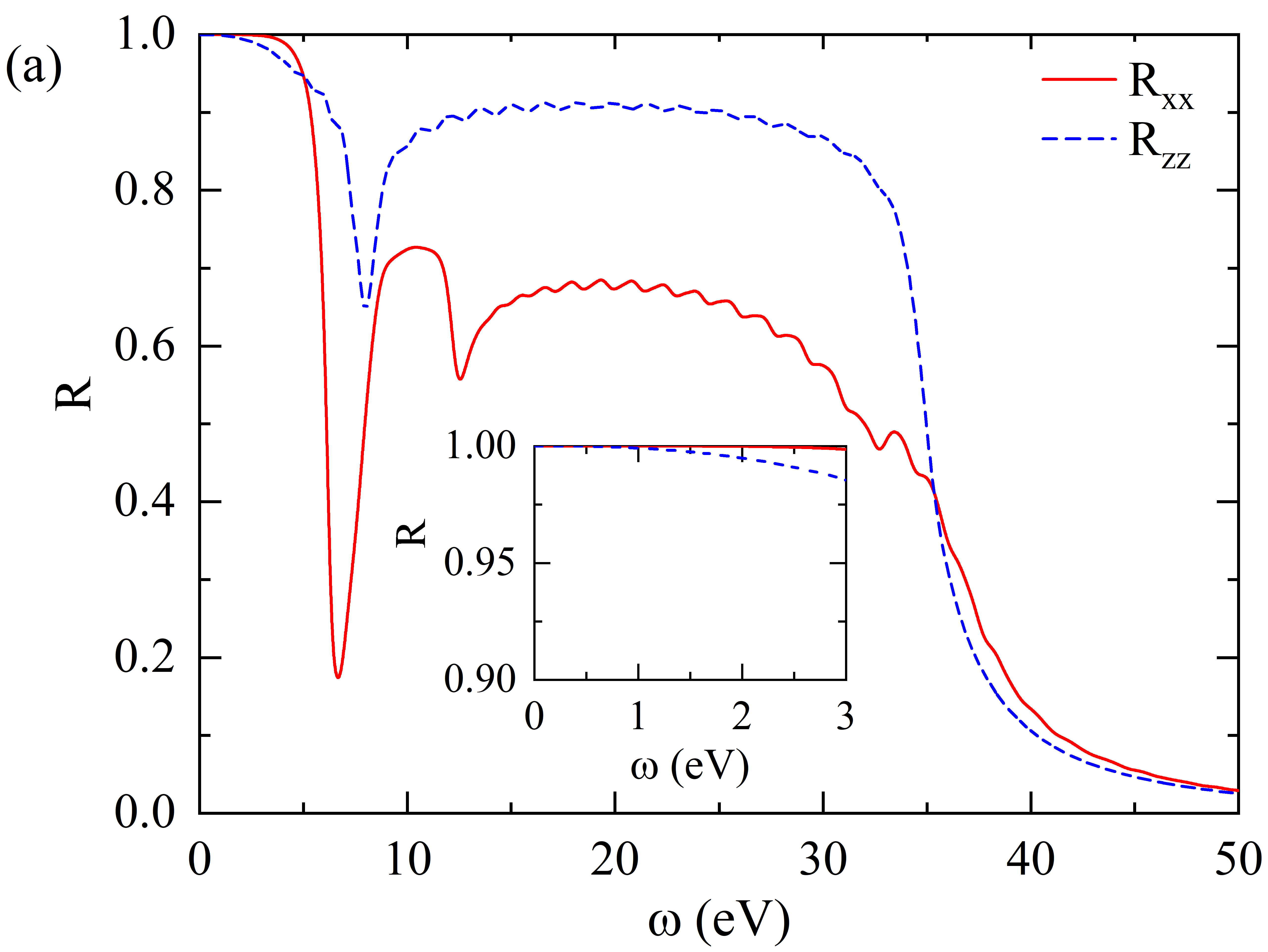}}
	\hspace{0.05cm}
	\subfloat{\includegraphics[width=0.395\textwidth]{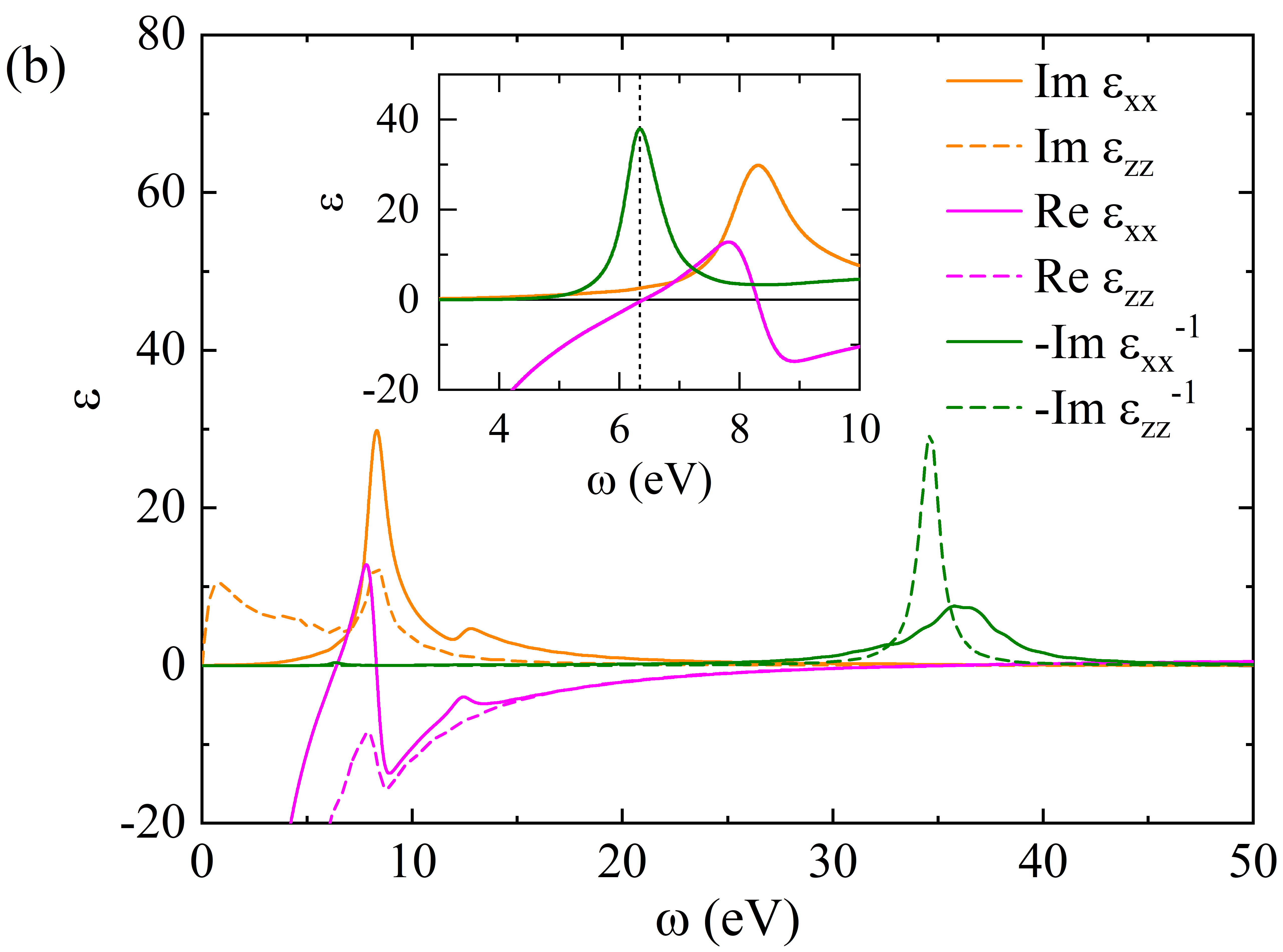}}
	\caption{\label{fig1}
	Static-nuclei optical spectra of I41/amd at 495 GPa. (a) the reflectance $R_{xx}$=$R_{yy}$ (red solid curve) and $R_{zz}$ (blue dash curve); (b) the real parts, $\text{Re}\,\varepsilon_{xx}=\text{Re}\,\varepsilon_{yy}$ (magenta solid curve) and $\text{Re} \,\varepsilon_{zz}$ (magenta dash curve), and imaginary parts, $\text{Im}\,\varepsilon_{xx}=\text{Im}\,\varepsilon_{yy}$ (orange solid curve) and $\text{Im}\,\varepsilon_{zz}$ (orange dash curve), of the dielectric functions as well as loss functions $\text{-Im}\,\varepsilon_{xx}^{-1}=\text{-Im}\,\varepsilon_{yy}^{-1}$ (olive solid curve) and $\text{-Im}\,\varepsilon_{zz}^{-1}$ (olive dash curve). $\text{Im}\,\varepsilon=\text{Im}\,\varepsilon_{\text{inter}}$ for $\omega>0$. The inset in (a) shows the reflectance in visible and IR ranges. The inset in (b) shows the interband  plasmon, where the loss function is multiplied by 100 and the dotted line labels the plasmon frequency (6.3 eV).
	}
\end{figure*}
We start with a general discussion on the relative stability of the candidate structures, since they
determine the electronic structures and consequently the optical properties accessible to the
experimental measurements in Ref.~\onlinecite{dias2017observation}.
This is done by resorting to published results from earlier theoretical studies.
Random structure searchings based on DFT show that a C2/c-24 phase is the most competitive
structure at $\sim$300 GPa,~\cite{Pickard2007,Li2013,drummond2015quantum} when the PBE functional is used.
Above 500 GPa, the molecular hydrogen will dissociate to an atomic
I41/amd phase.~\cite{mcmahon2011ground,McMinis2015,azadi2014dissociation}
Between 300 and 500 GPa, the existence of other molecular hydrogen phases is elusive, and the
Cmca-12 and Cmca-4 structures are competitive.~\cite{azadi2014dissociation}
Since the electronic structures from DFT using approximate functionals may be
inaccurate, higher level electronic structure methods were soon resorted to.
Diffusion quantum Monte Carlo (DMC) calculations indicate that Cmca-4 is unstable when including
zero-point motion (ZPM).~\cite{azadi2014dissociation}
Then, another quantum Monte Carlo (QMC) simulation shows that C2/C-24 is the most stable almost
till the transition to I41/amd.~\cite{McMinis2015}
Considering the fact that the enthalpy differences between these four structures are in the range
of a few meV/H and sensitive to the choice of electronic structure methods,~\cite{azadi2013fate,McMinis2015} we analyze the optical spectra for all of them when compared
with experiment, for a more realistic interpretation of the latter.\\
\indent
\indent
\begin{figure*}[ht!]
	\subfloat{\includegraphics[width=0.39\textwidth]{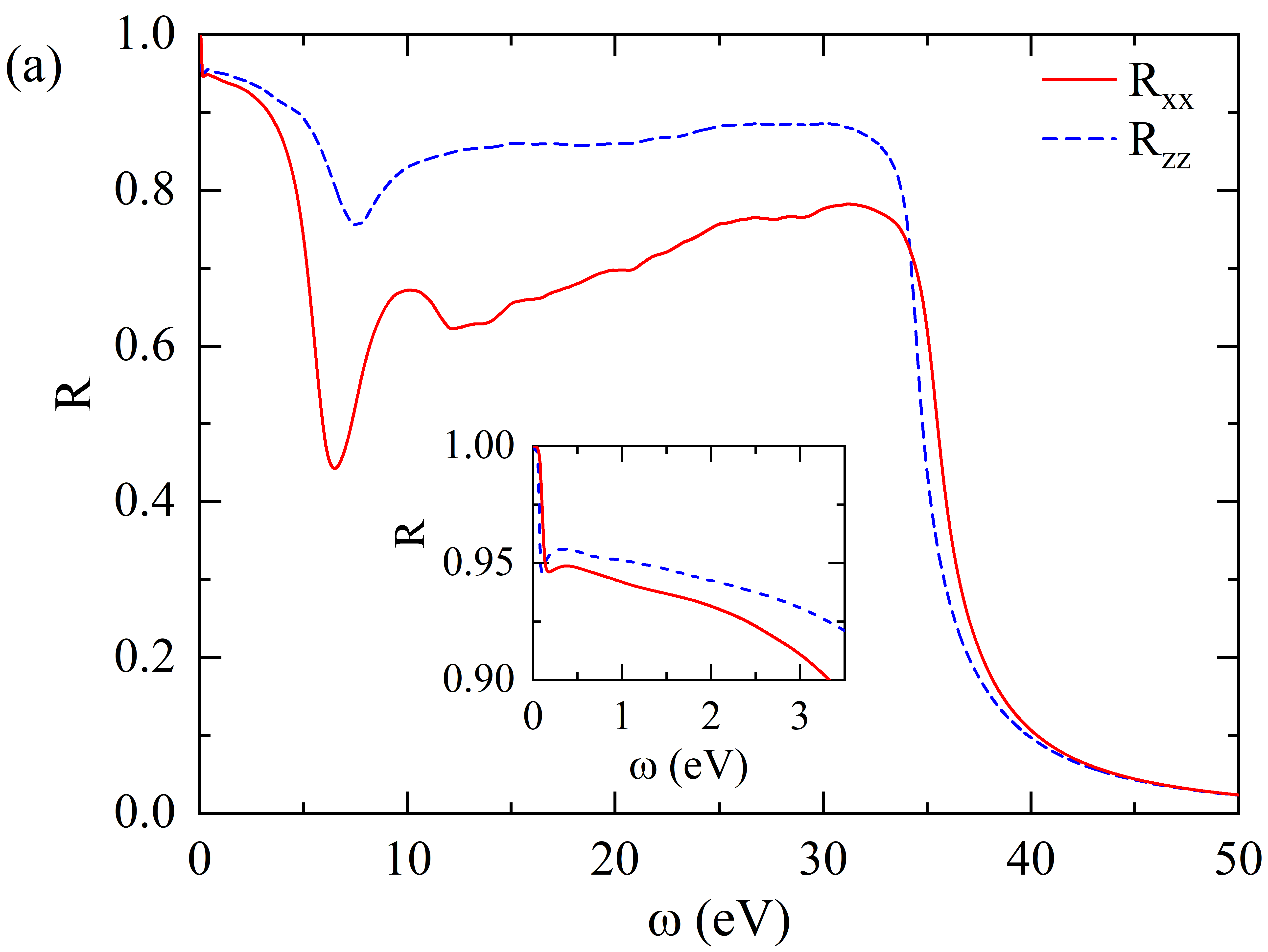}}
	\hspace{0.05cm}
	\subfloat{\includegraphics[width=0.39\textwidth]{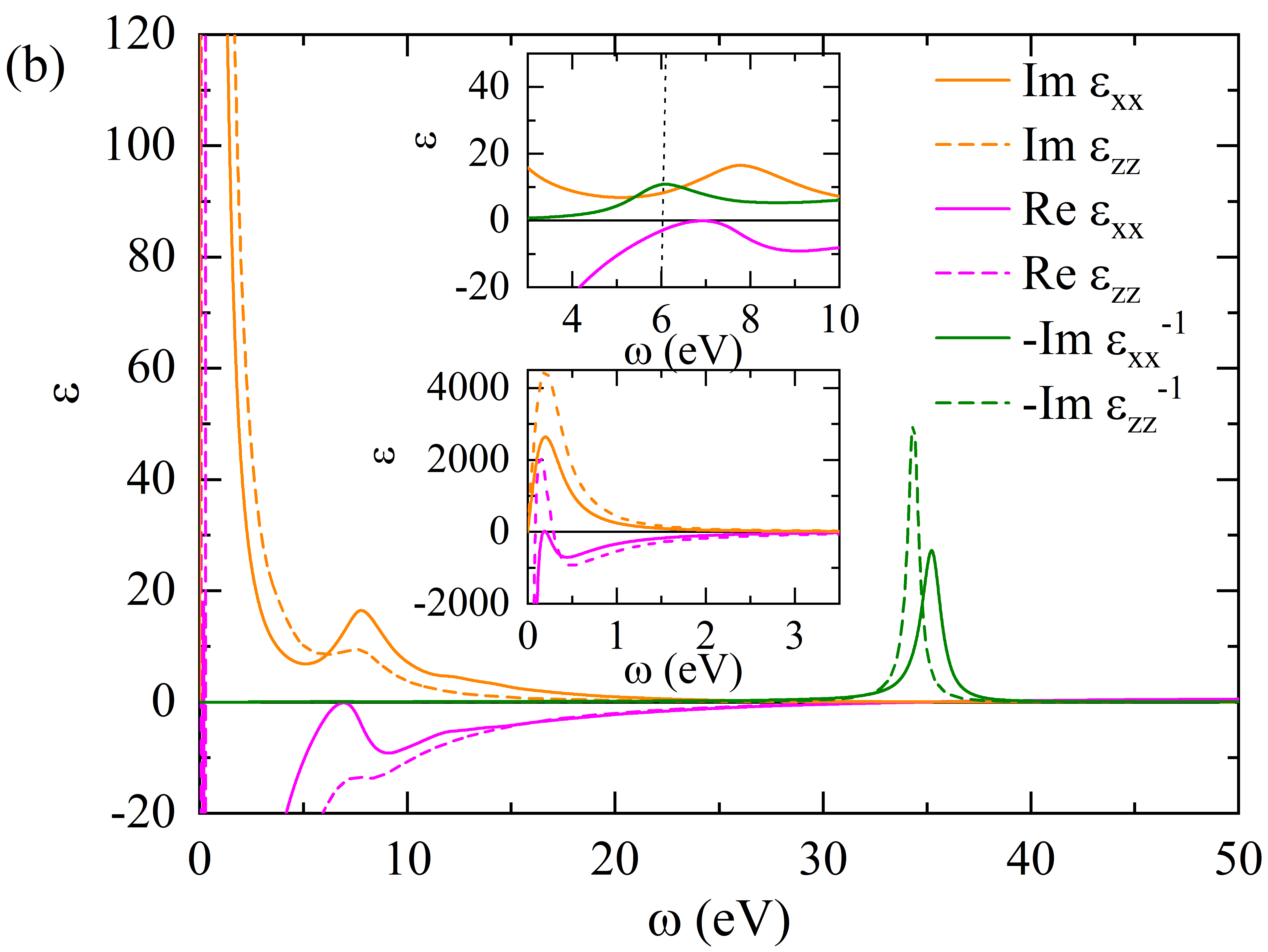}}
	\caption{\label{fig2}
	The optical spectra of I41/amd with EPIs using WL-HA at 495 GPa and 5 K. (a) the reflectance $R_{xx}$=$R_{yy}$ (red solid curve) and $R_{zz}$ (blue dash curve); (b) the real parts, $\text{Re}\,\varepsilon_{xx}=\text{Re}\,\varepsilon_{yy}$ (magenta solid curve) and $\text{Re} \,\varepsilon_{zz}$ (magenta dash curve), and imaginary parts, $\text{Im}\,\varepsilon_{xx}=\text{Im}\,\varepsilon_{yy}$ (orange solid curve) and $\text{Im}\,\varepsilon_{zz}$ (orange dash curve), of the dielectric functions as well as loss functions $\text{-Im}\,\varepsilon_{xx}^{-1}=\text{-Im}\,\varepsilon_{yy}^{-1}$ (olive solid curve) and $\text{-Im}\,\varepsilon_{zz}^{-1}$ (olive dash curve). $\text{Im}\,\varepsilon=\text{Im}\,\varepsilon_{\text{intra}}+\text{Im}\,\varepsilon_{\text{inter}}$. The inset in (a) shows the reflectance in visible and IR ranges. The upper inset in (b) shows the interband  plasmon, where the loss function is multiplied by 100 and the dotted line labels the plasmon frequency (6.1 eV). The bottom inset in (b) shows the dielectric functions in visible and IR ranges.
	}
\end{figure*}
A special focus of this work is to simulate the reflectance of H and compare it directly with
the DS's experiment.
An earlier theoretical study by Borinaga \textit{et al.} has shown that the EPIs are
important.~\cite{borinaga2018strong}
As such, we first present a comparison between the optical spectra of I41/amd, using the
static geometry optimized structure (without EPIs) and using the WL methods (with EPIs).
Fig.~\ref{fig1}a shows the reflectance of I41/amd with nuclei calmed at equilibrium
positions (without EPIs).
Since I41/amd belongs to the tetragonal crystal structure, symmetry requires the reflectance
to be $R_{xx}=R_{yy}\neq R_{zz}$.
They all have values close to 100$\%$ in the visible and infrared (IR) ranges
(0-3 eV, see the inset of Fig.~\ref{fig1}a).
Near 5 eV, both $R_{xx}$ ($R_{yy}$) and $R_{zz}$ decrease sharply, and after that they rapidly
rise back to relatively high values till 35 eV.
Concerning their differences due to anisotropy, $R_{zz}$ decreases faster than $R_{xx}(R_{yy})$ at low frequency, but the subsequent dip at 6-8 eV is clearly more serious in $R_{xx}$ ($R_{yy}$).
Specifically, the $R_{xx}(R_{yy})$ dip at 6.7 eV is deeper than the $R_{zz}$ dip at
7.8 eV.\\
\indent
\begin{figure*}
\centering
 \includegraphics[width=0.85\linewidth]{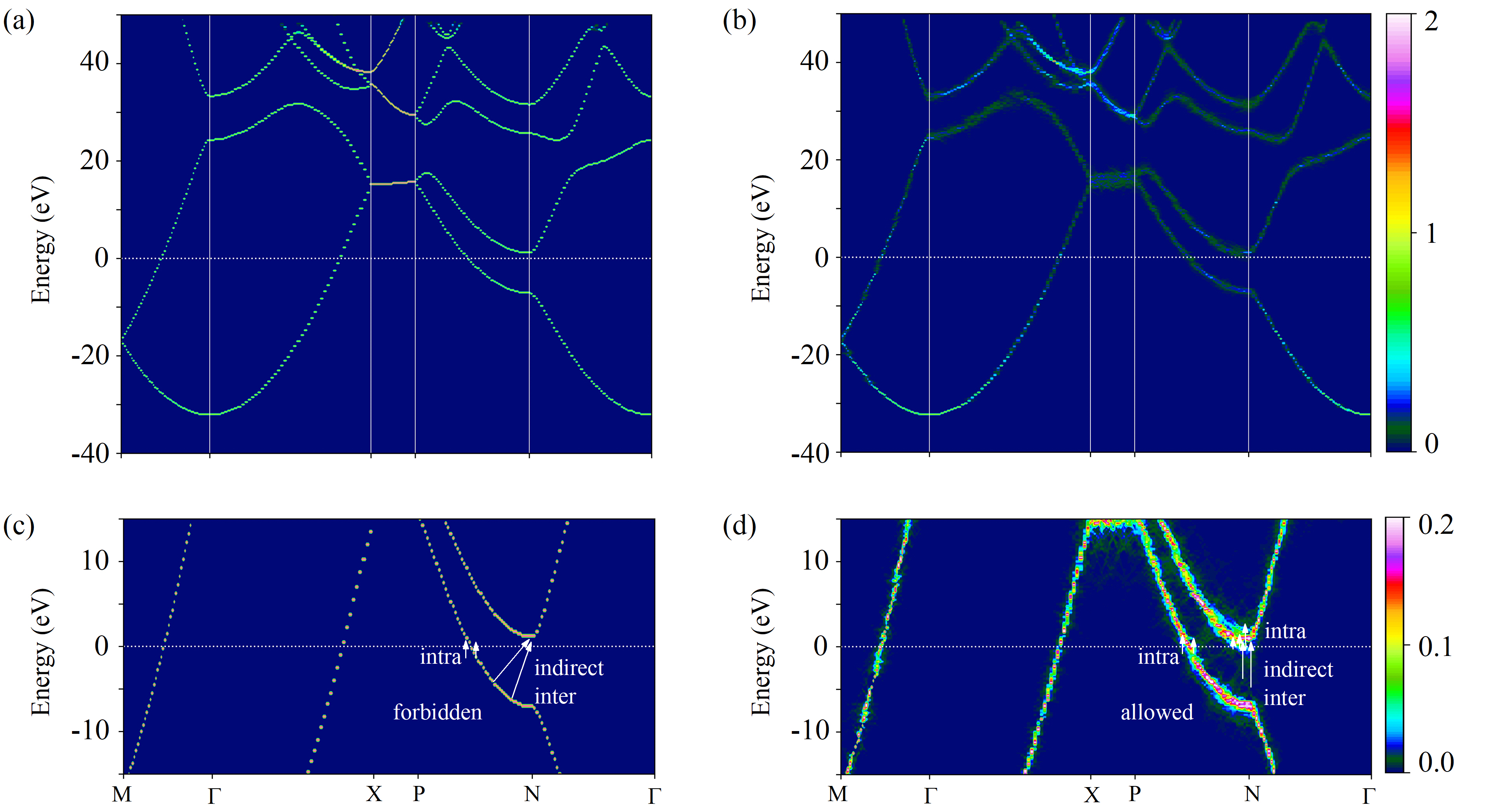}
 \caption{\label{fig3}Effective bandstructure of I41/amd at 495 GPa. (a) and (c) without EPIs; (b) and (d) with EPIs at 5 K using WL-HA. The white arrows in (c) and (d) highlight that those intraband and interband transitions forbidden in static-nuclei case are allowed when including EPIs. The ``intra'' means intraband transitions. The ``indirect inter" means indirect interband transitions and labels additional states appearing near N. }
\end{figure*}
These changes of the reflectance are closely
related to the imaginary and real parts of the dielectric functions, and the electronic energy
loss functions (imaginary part of the inverse dielectric function).
As such, we show these functions, i.e. Im~$\varepsilon$, Re~$\varepsilon$, and -Im~$\varepsilon^{-1}$,
in Fig.~\ref{fig1}b.
%
%
%
The reason of the sudden dip of the reflectance at 6-8 eV is associated with the peaks existing
in the Im~$\varepsilon$ and Re~$\varepsilon$ in the same frequency range.
In Sec. II A, we have shown that for metals the dielectric function consists of intraband and
interband contributions.
Examining the form of the intraband dielectric function in Eq.~\ref{realintra}, where
$\omega_{\text{p},zz}$=29.4 eV and $\omega_{\text{p},xx}$=22.6 eV, it is clear that
Re~$\varepsilon_{\text{intra}}$ (including Re~$\varepsilon_{\text{intra},xx}$ and Re~$\varepsilon_{\text{intra},zz}$) is structureless and it approaches 0 asymptotically as
the frequency increases.
Thus the peaks emerging in the Re~$\varepsilon$ at 6-8 eV (including Re~$\varepsilon_{xx}$
at 7.8 eV and Re~$\varepsilon_{zz}$ at 7.9 eV), which violate the asymptotic feature of Re~$\varepsilon_{\text{intra}}$, mean that the interband contributions begin to be
comparable to the intraband ones.
As a result of the Kramers-Kr\"onig relation, peaks will appear at subsequent frequencies in
Im~$\varepsilon$ (e.g. in Im~$\varepsilon_{xx}$ and Im~$\varepsilon_{zz}$ both at 8.3 eV)
due to interband transitions.
Before these peaks, i.e. in the range 0-5 eV, the intraband contribution to Re~$\varepsilon$
predominates.
During the range of these peaks (6-8 eV), the interband and intraband transitions have
comparable contributions.
\indent

For a direct analysis of how these changes in Im~$\varepsilon$ and Re~$\varepsilon$ impact
on the reflectance, we resort to Eqs.~\ref{refractive} to \ref{reflect}.
From these equations, it is clear that the reflectance is determined by
the comparison of the magnitudes of Im~$\varepsilon$ and Re~$\varepsilon$, and their absolute
values.
Below 5 eV, the magnitude of Im~$\varepsilon$ is much smaller that of Re~$\varepsilon$
and the absolute value of Re~$\varepsilon$ is orders of magnitude larger than 1.
From Eqs.~\ref{refractive} to \ref{reflect}, one can easily obtain a reflectance
close to 100$\%$.
During the range of the dip of the reflectance (6-8 eV), we have shown in the last paragraph
that the interband contributions to the real part of the dielectric function substantially
decrease the magnitude of Re~$\varepsilon$, making it comparable to that of Im~$\varepsilon$,
interesting phenomena appear.
The reflectance minimum in $R_{zz}$ (7.8 eV) is a consequence of the peak of
$\text{Re}\,\varepsilon_{zz}$ at 7.9 eV.
For $R_{xx}$ and $R_{yy}$, a more complicated scenario appears.
The interband transitions can result in an interband plasmon at 6.3 eV in
$-\text{Im}\,\varepsilon_{xx}^{-1}$, due to the fact that Re $\varepsilon_{xx}$ crosses zero at nearly
the same frequency.~\cite{borinaga2018strong}
At this point, the magnitude of Im $\varepsilon_{xx}$ dominates over Re $\varepsilon_{xx}$ but the absolute
value of Im $\varepsilon_{xx}$ is small, the reflectance suddenly dip from 1 when these values were
put into Eqs.~\ref{refractive} to \ref{reflect}.
This is shown in detail in the inset of Fig.~\ref{fig1}b, where at 6.3 eV -Im $\varepsilon_{xx}^{-1}$(-Im $\varepsilon_{yy}^{-1}$) has a sharp peak with small damping, i.e. small values of Im $\varepsilon_{xx}$
and Im $\varepsilon_{yy}$, and the real part of dielectric function crosses zero.
It is this weakly-damped plasmon originating from interband transitions that makes the reflectance
dip (6.7 eV) in $R_{xx}$ sharper than $R_{zz}$.
In addition to the low energy plasmon, there exist other plasma peaked at much higher energy (34.6 eV for
-Im $\varepsilon_{zz}^{-1}$, 35.8 eV for -Im $\varepsilon_{xx}$), where both the real and the imaginary
part of the dielectric functions approach 0.
These plasma are responsible for the final decrease of the reflectance over 35 eV and are called
free-electron plasma with the plasmon frequencies being close to the theoretical value, i.e. $\omega_{\text{p}}=\sqrt{4\pi n e^2/m}=34.7$ eV, where $n$ is the electron density and $m$ is the
electron mass.\\
\indent
Our above DFT-IPA results for $R_{xx}$ are consistent with the TDDFT ones in Ref.~\onlinecite{borinaga2018strong}, which justified that for metals IPA is a good approximation due
to the cancellation of errors originating from neglecting the electron-electron interactions and
the electron-hole interactions.~\cite{marini2001optical}
These two interactions are purely between electrons.
Concerning EPIs, in Ref~\onlinecite{borinaga2018strong}, only intraband transitions were considered
in solving the isotropic Migdal-Eliashberg equation.
In the following, we do two major extensions, i.e. i) using WL-HA to investigate the influences of EPIs
on the reflectance with both interband and interband transitions included, and ii) addressing
anisotropy.\\
\indent
In Fig.~\ref{fig2}a, we present the reflectance of I41/amd with EPIs included.
With WL-HA, the reflectance is independent of $T$ below 83~K.
The results shown are for 5~K, and these curves don't change for other $T\le83$~K.
From Fig.~\ref{fig2}a, we see that the reflectance has some noticeable changes compared with
the static-nuclei one.
The most apparent two are: i) in the visible and IR regions the reflectance decreases to
below 95$\%$ (see the inset of Fig.~\ref{fig2}a), and ii) the dips at 7.6 eV (6.5 eV) for
$R_{zz}$ ($R_{xx}$) become weaker and broader and they have red shift of $\sim$0.2 eV.
Again, these changes can be explained by the dielectric functions and the loss functions with EPIs,
as shown in Fig.~\ref{fig2}b.
Below 5 eV, there are two orders of magnitude increase in Im $\varepsilon$ (see the bottom inset of Fig.~\ref{fig2}b), due to the fact that our EPIs treatment has effectively included the intraband
transitions.
To be specific, the occupation numbers were smeared and a finite electron lifetime is induced due to
EPIs.
In so doing, the intraband transitions are allowed.
This is also shown in Fig.~\ref{fig3}d, and we will explain later.
We note that without EPIs (clamped structure) the Im $\varepsilon_{\text{intra}}$ is rigorously zero
for nonzero frequencies within IPA and the total Im $\varepsilon$ is small.
The reflectance is nearly 100$\%$ at low frequencies due to the large magnitude of Re $\varepsilon$.
With EPIs, it is the comparable values of Im $\varepsilon$ and Re $\varepsilon$
resulting from the intraband transition, which induce the drop of reflectance from nearly $100\%$
to below 95$\%$ at low frequencies.
At higher energies, the significantly weakening of the reflectance dips in $R_{xx}$ and $R_{zz}$ is
closely related to the weakening of the peaks in Im $\varepsilon$ and
Re $\varepsilon$ (Fig.~\ref{fig2}b).
This is most obvious in $R_{xx}$ and the smearing of the interband plasmon peak (reflected by
the loss function) at $\sim$6.1 eV plays a crucial role.
We show this in detail in the upper inset of Fig.~\ref{fig2}b.
The peaks of Im $\varepsilon_{xx}$ and Re $\varepsilon_{zz}$ are much broader compared with their
static clamped nuclei correspondences in the inset of Fig.~\ref{fig1}b.
The peak of the loss function is also much weaker.\\
\indent
The above analysis shows that the EPIs play an important role in the reflectance by
modulating the contributions from the intraband and interband transitions.
To elucidate this point more clearly, we present the band structures of I41/amd with and without
EPIs by using band unfolding method,~\cite{medeiros2014effects,medeiros2015unfolding} and show the
results in Fig.~\ref{fig3}.
With static nuclei (Fig.~\ref{fig3}a), the color intensity is 1 or 2, which is the band
degeneracy (2 is the highest allowed degenerate number with this symmetry).
With EPIs (Fig.~\ref{fig3}b), lattice distortion results in the color intensity mostly with fractional
numbers, except for the deep states.
Considering the fact that during the EPIs, the phonon contributes momentum to the electronic
states and the total energy is also required to be conserved, the band with large dispersion
will often be smeared less than those flat ones.
This is most apparent if we compare the lowest flat band between X and P with the lowest parabolic
band between $\Gamma$ and X.
For the flat band, the electronic states at neighboring $\mathbf{k}$-points have dispersions.
Phonon can scatter electrons from these dispersed states to the flat states when EPIs are included.
In so doing, they are smeared.\\
\indent
These changes of the occupation numbers significantly influence the optical transitions by the
Fermi's Golden rule.
This is illustrated by the white arrows in Figs.~\ref{fig3}c and~\ref{fig3}d.
With static nuclei, the intraband transitions are forbidden (Fig.~\ref{fig3}c).
Indirect interband transitions indicated by inclined arrows are not allowed either, due to the
conservation of momentum in Eq.~\ref{imagepsilon}.
With EPIs, however, both these processes are can happen (Fig.~\ref{fig3}d).
For the intraband transitions, since the band becomes broad near the Fermi surface with fractional
occupation number, an electron can easily jump from an occupied state to an unoccupied one within the
same band.
These transitions around the Fermi surface contribute a large part of the increase in the imaginary
part of the dielectric functions below 5 eV in Fig.~\ref{fig2}b.
In addition to these intraband transitions, the EPIs may also induce additional states at a
certain $\mathbf{k}$-point, originating from electronic states at neighboring $\mathbf{k}$-points.
In so doing, the momentum is conserved during the interband optical transition, and we label such
processes as ``indirect-inter'' in Fig.~\ref{fig3}d.

\subsection{Reflectance}
\indent
\begin{figure*}[ht!]
	\subfloat{\includegraphics[width=0.41\textwidth]{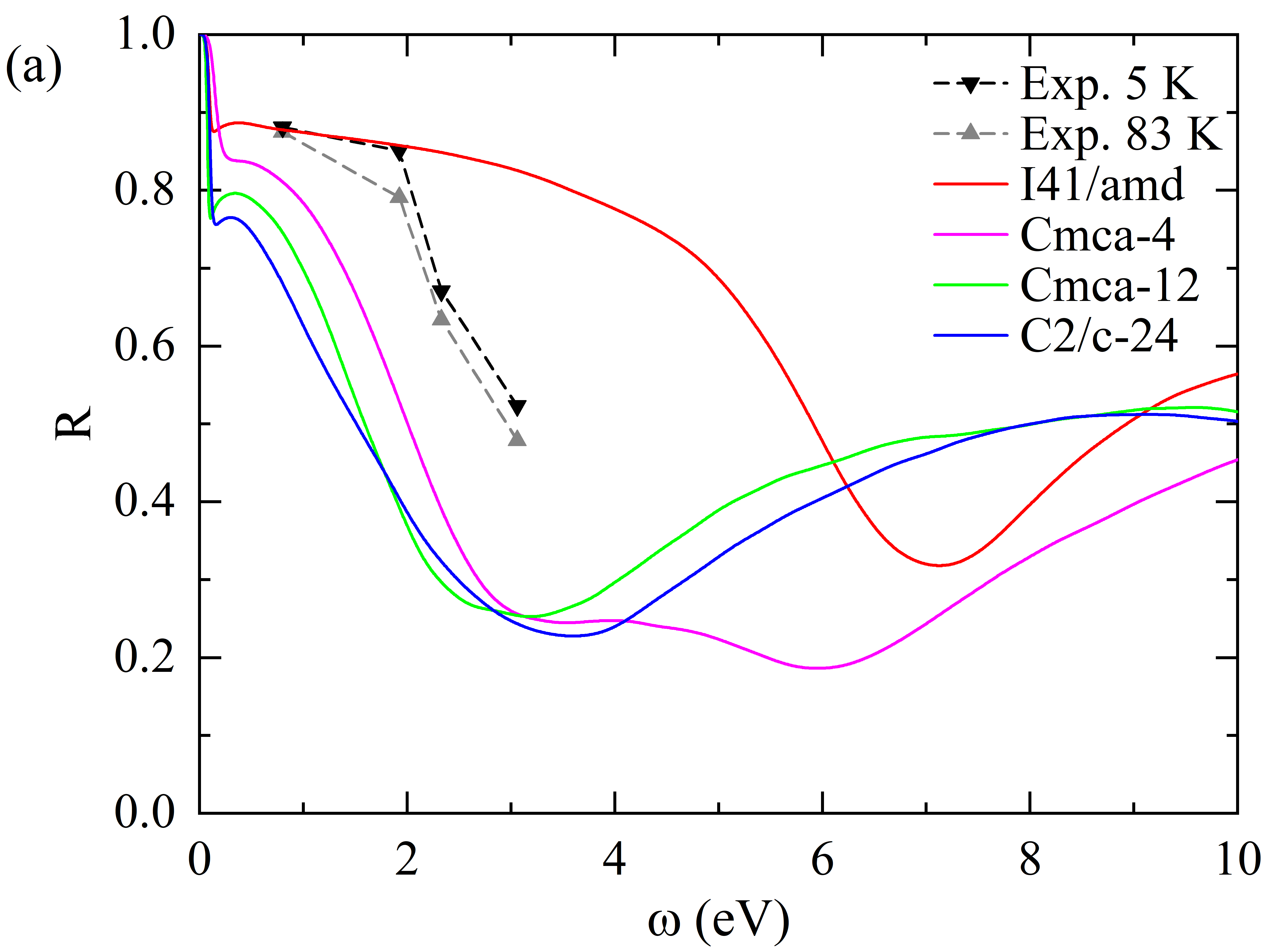}}
	\hspace{0.05cm}
	\subfloat{\includegraphics[width=0.41\textwidth]{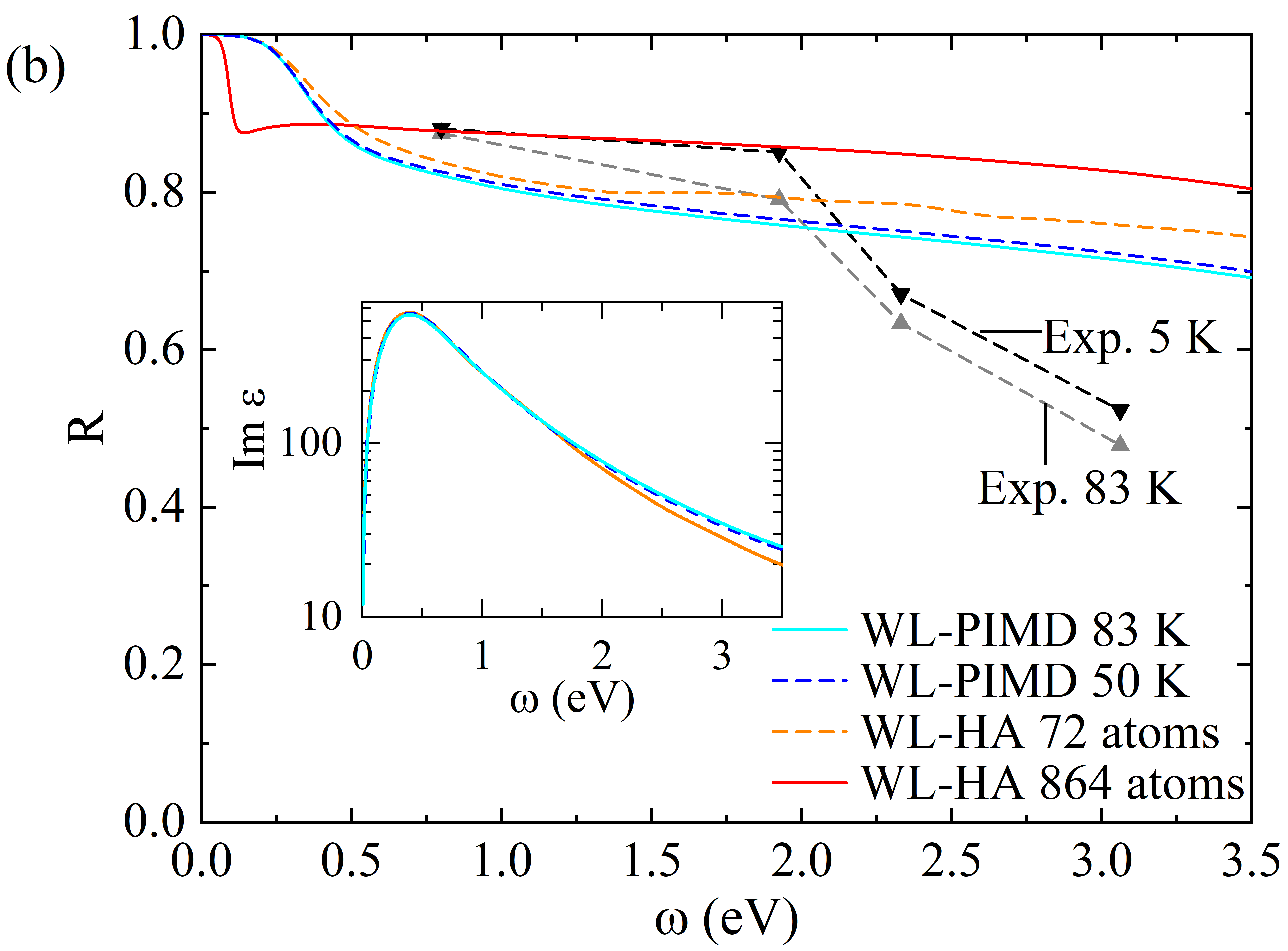}}
	\caption{\label{fig4}
	(a) Comparisons between DS's experiment at 5 K (black lower-triangle dash line) and 83 K (gray upper-triangle dash line) with the pressure being 495 GPa and the WL-HA diamond/hydrogen interface reflectance of four structures at 495 GPa and 5 K, i.e. C2/c-24 (blue solid curve), Cmca-12 (green solid curve), Cmca-4 (magenta solid curve) and I41/amd (red solid curve). (b) The influences of nuclear AHEs on the reflectance of I41/amd at 495 GPa by comparing the results of WL-HA at 5 K using 72-atom (orange dash curve) and 864 atom (red solid curve) unit cell and WL-PIMD at 50 K (blue dash curve) and 83 K (cyan solid curve) using 72-atom unit cell. The inset of (b) shows the imaginary parts ($\text{Im}\,\varepsilon=\text{Im}\,\varepsilon_{\text{intra}}+\text{Im}\,\varepsilon_{\text{inter}}$) of the dielectric funtions within 72-atom unit cell by using WL-HA (orange solid curve) and WL-PIMD 50 K (blue dash curve) and 83 K (cyan solid curve). The diamond's refractive index is 2.41 for the calculations of diamond/hydrogen interface reflectance  in (a) and (b).
	}
\end{figure*}
\begin{figure*}[ht!]
	\subfloat{\includegraphics[width=0.40\textwidth]{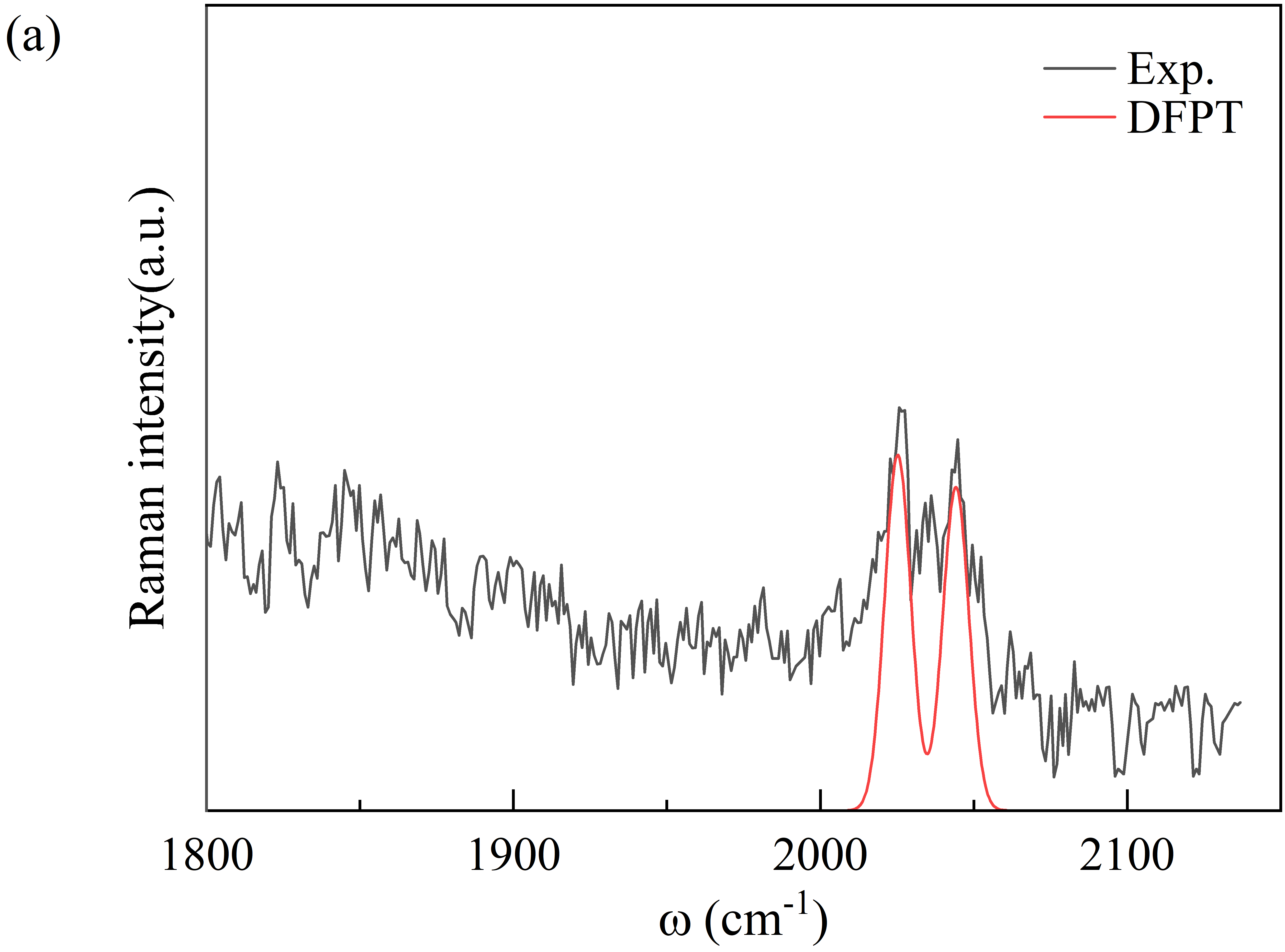}}
	\hspace{0.05cm}
	\subfloat{\includegraphics[width=0.41\textwidth]{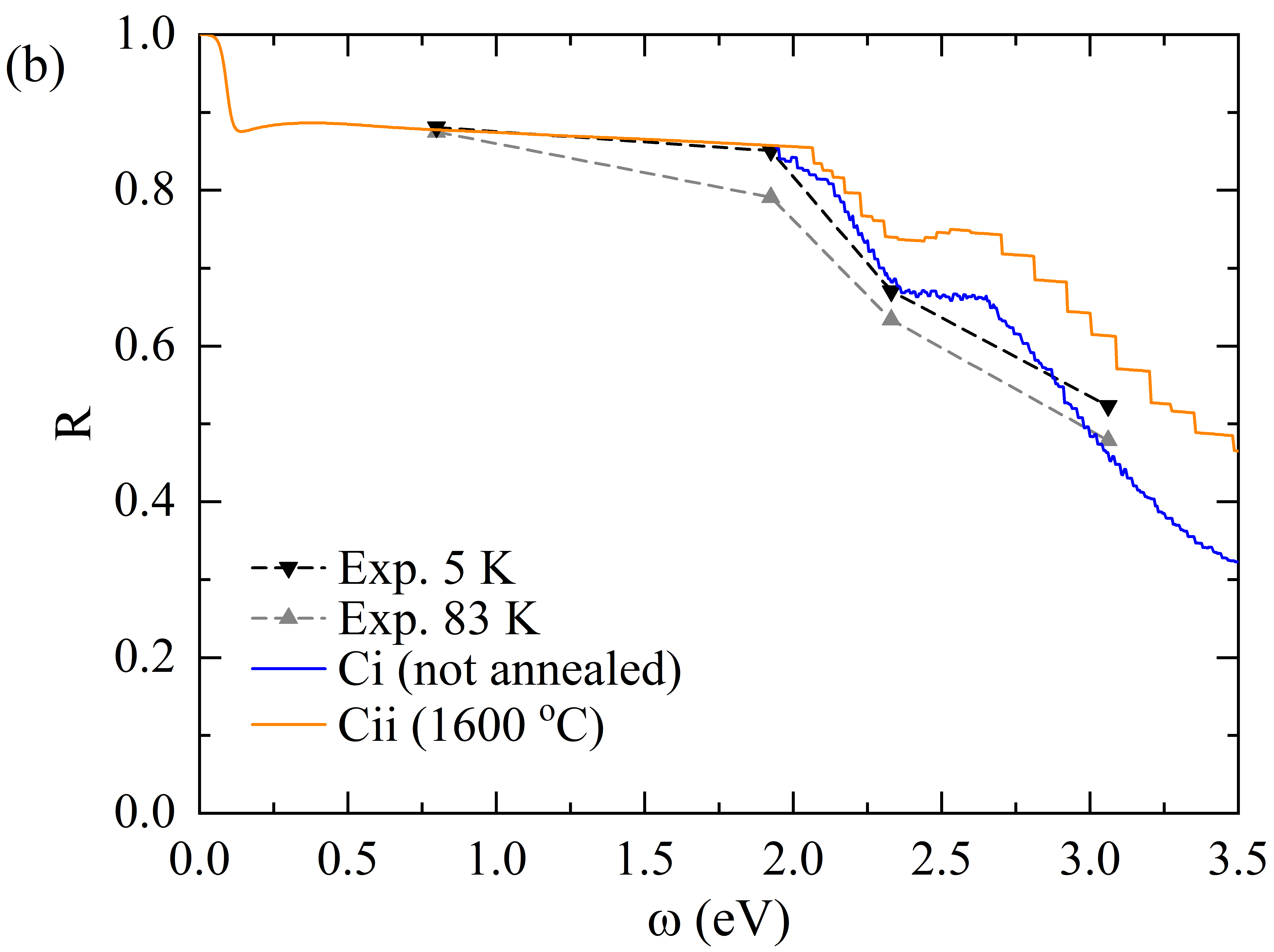}}
	\caption{\label{fig5}
		a) Comparisons between calculated tetragonal diamond's Raman spectrum using DFPT (red solid curve) and DS's experiment (black solid curve) at 495 GPa. b) Comparisons between DS's experiment at 5 K (black lower-triangle dash line) and 83 K (gray upper-triangle dash line) with the pressure being 495 GPa and corrected WL-HA diamond/hydrogen interface reflectance of I41/amd at 495 GPa and 5 K. Ci (blue solid curve) and Cii (orange solid curve) are diamond samples not annealed and annealed at 1600$^{\circ}$C, respectively. The diamond's refractive index is 2.41 for the calculations of diamond/hydrogen interface reflectance.
	}
\end{figure*}
With the concepts about EPIs discussed above, we now look at the four most competitive MH structures,
i.e. C2/c-24, Cmca-12, Cmca-4 and I41/amd, and compare their reflectance directly with DS's experiment.
The four structures are not cubic and they all have anisotropic optical properties.
In DS's experiment, the relation between the incident light's polarization direction and the
crystal structure is unknown, and the sample is most likely polycrystalline.
Therefore, we average the diagonal terms of the theoretical dielectric tensors when comparing with
experiment.
In Fig.~\ref{fig4}a, we present the diamond/hydrogen interface reflectance of these four structures
using WL-HA.
It can be seen that the C2/c-24, Cmca-12, and Cmca-4 structures have similar reflectance and
the drop happens at much lower energy than I41/amd.
The deviations of the former three structures from experiment are obvious.
The reflectance of I41/amd, however, agrees well with the experiment below 2 eV.
This comparison supports that I41/amd is the most possible MH candidate for the DS's experiment.
But we note that two experimental features, i.e. the large $T$-dependence and the drop above 2 eV of
the reflectance, are still unexplained.\\
\indent
To explore the $T$-dependence of the experimental reflectance, we further consider nuclear AHEs,
stimulated by the result in Ref.~\onlinecite{azadi2014dissociation} that AHEs induce more
delocalized nuclei.
This is done by comparing the WL-HA results with the WL-PIMD ones, as shown in Fig.~\ref{fig4}b.
To avoid the inaccuracy originating from finite-size effects, we first compare the reflectance obtained
using the WL-HA and WL-PIMD methods with the same supercell (i.e. 72 atoms).
Below 1.5 eV, the differences between the WL-HA and the WL-PIMD results are small.
Above this value, their differences become apparent because AHEs can help interband transitions
(see the inset of Fig.~\ref{fig4}b).
It is worth noting that below 3.5 eV the WL-HA results with 72-atom supercell are lower than the WL-HA results with 864-atom supercell, meaning that the WL-HA and the WL-PIMD results with 72-atom supercell
both have large finite-size errors.
We use the difference between the WL-HA results with 72-atom and 864-atom supercells to estimate and
to correct this finite-size error.
After correction, it is fair to say that: i) the WL-PIMD results can present a good estimation
of the experimental reflectance, ii) the AHEs are non-negligible, and iii) the $T$-dependence absent
within WL-HA becomes appreciable when AHEs were taken into account.
However, this $T$-dependence is still not comparable to DS's experimental observation.
Considering the fact that the reflectance of Re gasket in DS's experiment assumes similar
$T$-dependence as that of H, it is likely that this large $T$-dependence is not intrinsic in
MH and it could be caused by other external reasons.\\
\indent
In the end, we look at the reflectance drop above 2 eV.
In Ref.~\onlinecite{borinaga2018strong}, this drop is speculated to be associated with the
interband plasmon.
However, as we have shown in Sec. III A, the interband plasmon is around 6 eV and it can not be
responsible for this drop at 2 eV.
Another reason is the diamond's gap reducing under anisotropic compression at high pressure,
as proposed in the original DS's paper.~\cite{dias2017observation}
To clarify this, we simulate the diamond's anisotropic compression by using the tetragonal diamond model proposed in Ref.~\onlinecite{surh1992band}.
The $c$/$a$ ratio and the diamond's lattice constants are obtained by matching the calculated Raman
spectra with the experiment in Ref.~\onlinecite{dias2017observation}.
As shown in Fig.~\ref{fig5}a, the calculated Raman spectra can match with the experiment.
When this happens, the pressures corresponding to these lattice constants are 430 GPa
along $a$ and $b$ axis, and 530 GPa along $c$.
This is close to the pressure (495 GPa) claimed in Ref.~\onlinecite{dias2017observation}.
Using this structure, we calculate the band gap using the $G_{0}W_{0}$ method based on
LDA.~\cite{hedin1965, hybertsen1986, shishkin2006, shishkin2007, ricardo2008, li2012, jiang2013}
The result is 4.05 eV.
This value is well above 2 eV, meaning that the drop at 2 eV is not caused by the pressure-induced
band gap reduction either. \\
\indent
The last possible reason proposed in earlier literature is the defects
in diamonds.~\cite{silvera2017response}
To address this, we resort directly to the absorption experiments and correct our reflectance
curve using their results.
The diamond used in the DS's experiment is type IIac, with ``c'' meaning chemical vapor deposition (CVD).
Therefore, we focus on the experiments for the CVD-grown type IIa diamond.
The absorption spectra below the band gap is mainly composed of three parts: 520 nm (2.39 eV) band, 360 nm (3.49 eV) band, and a featureless profile.~\cite{khan2013colour}
When annealed, the 520 nm band is removed at 1800 $^{\circ}$C, while the other two are significantly reduced at 1600 $^{\circ}$C.~\cite{khan2013colour}
In DS's experiment, the annealing $T$ is 1200 $^{\circ}$C, well below 1800 $^{\circ}$C.
Therefore, the 2.39 eV band should remain.
We choose two absorption curves from the experiment of Khan \textit{et al.} which include about 1 ppm
nitrogen defects resembling the experiments of DS,~\cite{khan2013colour} and correct the calculated diamond/hydrogen reflectance using:
\begin{equation}
R_{1}=R_{0}e^{-2\alpha\l}.
\end{equation}
Here $R_{0}$ ($R_{1}$) denotes uncorrected (corrected) reflectance, $\alpha$ is the absorption coefficient of diamonds and $\l\sim2\,\text{mm}$ is the diamond height in DS's experiment.
In Fig.~\ref{fig5}b, we show the results obtained after this correction.
For the not annealed case, the corrected reflectance is in good agreement with the experiment, although slightly lower than the data at 3.06 eV.
For annealed case, the corrected reflectance is higher than the experiment, but there is still a
significant drop.
Based on this, we expect that the absorption of the defects in diamond should be responsible for the
sudden drop of the reflectance at 2 eV.
\section{Conclusion}
\indent
As the \textit{Holly Grail} in high pressure physics, the experimental verification of MH is
challenging.
Existing reports can easily be controversial due to some prominent technical difficulties
in calibrating the pressure, and the indirect nature of the characterization of the crystal
and electronic structures.
In depth understanding of the available experiment data from the theoretical perspective, therefore,
is highly desired.\\
\indent
We present in this paper such an analysis on the optical spectra of MH, close to the pressure
range of the DS's experiment.
Special focus was put on the role of EPIs and on comparisons of the reflectance directly with
experiments.
Four candidate structures, i.e. C2/c-24, Cmca-12, Cmca-4, and I41/amd, were chosen.
These structures were thought to be the most competitive H structures at the DS's experiment claimed
pressure range ($\sim$500 GPa) in terms of static enthalpy, and when the ZPE corrections were included.
We found that the atomic I41/amd phase can result in reflectance in good agreement with the DS's
experiment, and the EPIs play an important role.
The reflectance curves of all the other three structures, on the other hand, are much worse.
Besides this, we also found that the AHEs, effects often left out in other theoretical studies of
the EPIs, were non-negligible.
These effects, however, are not sufficient to account for the $T$-dependence of the experimental
observed reflectance.
Therefore, this $T$-dependence should not be intrinsic to MH.
Concerning the drop of the reflectance at 2 eV, our calculations clearly show that it is not
caused by the diamond's band gap reducing or the interband plasmon.
Rather, the diamond's defects absorption is very likely to be main reason, since correcting our
calculated reflectance using experimental absorption data of diamond's defects can reproduce
the reflectance drop above 2 eV.
These results provide theoretical supports for the recent DS's experimental realization
of MH.
Our analysis of the EPIs also indicates that the static treatment of the nuclei is
far from being enough in describing such optical and electronic structures.
We highly recommend quantum treatments of both the electrons and nuclei with AHEs taken into
account in future studies.\\
\begin{acknowledgements}
X.W.Z., E.G.W. and X.Z.L. are supported by the National Key R \&D Program under Grant
Nos. 2016YFA0300901 and 2017YFA0205000, and NSFC (11774004, 11274012, 91021007, 11634001).
We would like to thank Professor R. P. Dias and Professor I. F. Silvera for helpful replys. I appreciate valuable suggestions from my colleagues, Q. J. Ye, W. Fang, D. Kang, X. F. Zhang and T. Shen. We are grateful for computational resources provided by TianHe-1A in Tianjin, China and Weiming No.1 at Peking University, Beijing, China.
\end{acknowledgements}


\begin{thebibliography}{94}%
\makeatletter
\providecommand \@ifxundefined [1]{%
 \@ifx{#1\undefined}
}%
\providecommand \@ifnum [1]{%
 \ifnum #1\expandafter \@firstoftwo
 \else \expandafter \@secondoftwo
 \fi
}%
\providecommand \@ifx [1]{%
 \ifx #1\expandafter \@firstoftwo
 \else \expandafter \@secondoftwo
 \fi
}%
\providecommand \natexlab [1]{#1}%
\providecommand \enquote  [1]{``#1''}%
\providecommand \bibnamefont  [1]{#1}%
\providecommand \bibfnamefont [1]{#1}%
\providecommand \citenamefont [1]{#1}%
\providecommand \href@noop [0]{\@secondoftwo}%
\providecommand \href [0]{\begingroup \@sanitize@url \@href}%
\providecommand \@href[1]{\@@startlink{#1}\@@href}%
\providecommand \@@href[1]{\endgroup#1\@@endlink}%
\providecommand \@sanitize@url [0]{\catcode `\\12\catcode `\$12\catcode
  `\&12\catcode `\#12\catcode `\^12\catcode `\_12\catcode `\%12\relax}%
\providecommand \@@startlink[1]{}%
\providecommand \@@endlink[0]{}%
\providecommand \url  [0]{\begingroup\@sanitize@url \@url }%
\providecommand \@url [1]{\endgroup\@href {#1}{\urlprefix }}%
\providecommand \urlprefix  [0]{URL }%
\providecommand \Eprint [0]{\href }%
\providecommand \doibase [0]{http://dx.doi.org/}%
\providecommand \selectlanguage [0]{\@gobble}%
\providecommand \bibinfo  [0]{\@secondoftwo}%
\providecommand \bibfield  [0]{\@secondoftwo}%
\providecommand \translation [1]{[#1]}%
\providecommand \BibitemOpen [0]{}%
\providecommand \bibitemStop [0]{}%
\providecommand \bibitemNoStop [0]{.\EOS\space}%
\providecommand \EOS [0]{\spacefactor3000\relax}%
\providecommand \BibitemShut  [1]{\csname bibitem#1\endcsname}%
\let\auto@bib@innerbib\@empty
\bibitem [{\citenamefont {Wigner}\ and\ \citenamefont
  {Huntington}(1935)}]{wigner1935possibility}%
  \BibitemOpen
  \bibfield  {author} {\bibinfo {author} {\bibfnamefont {E.}~\bibnamefont
  {Wigner}}\ and\ \bibinfo {author} {\bibfnamefont {H.~B.}\ \bibnamefont
  {Huntington}},\ }\href {\doibase 10.1063/1.1749590} {\bibfield  {journal}
  {\bibinfo  {journal} {J.Chem.Phys.}\ }\textbf {\bibinfo {volume} {3}},\
  \bibinfo {pages} {764} (\bibinfo {year} {1935})}\BibitemShut {NoStop}%
\bibitem [{\citenamefont {Ashcroft}(1968)}]{ashcroft1968metallic}%
  \BibitemOpen
  \bibfield  {author} {\bibinfo {author} {\bibfnamefont {N.~W.}\ \bibnamefont
  {Ashcroft}},\ }\href {\doibase 10.1103/PhysRevLett.21.1748} {\bibfield
  {journal} {\bibinfo  {journal} {Phys.Rev.Lett.}\ }\textbf {\bibinfo {volume}
  {21}},\ \bibinfo {pages} {1748} (\bibinfo {year} {1968})}\BibitemShut
  {NoStop}%
\bibitem [{\citenamefont {Loubeyre}\ \emph {et~al.}(1996)\citenamefont
  {Loubeyre}, \citenamefont {LeToullec}, \citenamefont {Hausermann},
  \citenamefont {Hanfland}, \citenamefont {Hemley}, \citenamefont {Mao},\ and\
  \citenamefont {Finger}}]{loubeyre1996x}%
  \BibitemOpen
  \bibfield  {author} {\bibinfo {author} {\bibfnamefont {P.}~\bibnamefont
  {Loubeyre}}, \bibinfo {author} {\bibfnamefont {R.}~\bibnamefont {LeToullec}},
  \bibinfo {author} {\bibfnamefont {D.}~\bibnamefont {Hausermann}}, \bibinfo
  {author} {\bibfnamefont {M.}~\bibnamefont {Hanfland}}, \bibinfo {author}
  {\bibfnamefont {R.~J.}\ \bibnamefont {Hemley}}, \bibinfo {author}
  {\bibfnamefont {H.~K.}\ \bibnamefont {Mao}}, \ and\ \bibinfo {author}
  {\bibfnamefont {L.~W.}\ \bibnamefont {Finger}},\ }\href {\doibase
  10.1038/383702a0} {\bibfield  {journal} {\bibinfo  {journal} {Nature}\
  }\textbf {\bibinfo {volume} {383}},\ \bibinfo {pages} {702} (\bibinfo {year}
  {1996})}\BibitemShut {NoStop}%
\bibitem [{\citenamefont {Narayana}\ \emph {et~al.}(1998)\citenamefont
  {Narayana}, \citenamefont {Luo}, \citenamefont {Orloff},\ and\ \citenamefont
  {Ruoff}}]{narayana1998solid}%
  \BibitemOpen
  \bibfield  {author} {\bibinfo {author} {\bibfnamefont {C.}~\bibnamefont
  {Narayana}}, \bibinfo {author} {\bibfnamefont {H.}~\bibnamefont {Luo}},
  \bibinfo {author} {\bibfnamefont {J.}~\bibnamefont {Orloff}}, \ and\ \bibinfo
  {author} {\bibfnamefont {A.~L.}\ \bibnamefont {Ruoff}},\ }\href {\doibase
  10.1038/29949} {\bibfield  {journal} {\bibinfo  {journal} {Nature}\ }\textbf
  {\bibinfo {volume} {393}},\ \bibinfo {pages} {46} (\bibinfo {year}
  {1998})}\BibitemShut {NoStop}%
\bibitem [{\citenamefont {Goncharov}\ \emph {et~al.}(2001)\citenamefont
  {Goncharov}, \citenamefont {Gregoryanz}, \citenamefont {Hemley},\ and\
  \citenamefont {Mao}}]{goncharov2001spectroscopic}%
  \BibitemOpen
  \bibfield  {author} {\bibinfo {author} {\bibfnamefont {A.~F.}\ \bibnamefont
  {Goncharov}}, \bibinfo {author} {\bibfnamefont {E.}~\bibnamefont
  {Gregoryanz}}, \bibinfo {author} {\bibfnamefont {R.~J.}\ \bibnamefont
  {Hemley}}, \ and\ \bibinfo {author} {\bibfnamefont {H.~K.}\ \bibnamefont
  {Mao}},\ }\href {\doibase 10.1073/pnas.201528198} {\bibfield  {journal}
  {\bibinfo  {journal} {Proc. Natl. Acad. Sci. U.S.A}\ }\textbf {\bibinfo
  {volume} {98}},\ \bibinfo {pages} {14234} (\bibinfo {year}
  {2001})}\BibitemShut {NoStop}%
\bibitem [{\citenamefont {Bonev}\ \emph {et~al.}(2004)\citenamefont {Bonev},
  \citenamefont {Schwegler}, \citenamefont {Ogitsu},\ and\ \citenamefont
  {Galli}}]{bonev2004quantum}%
  \BibitemOpen
  \bibfield  {author} {\bibinfo {author} {\bibfnamefont {S.~A.}\ \bibnamefont
  {Bonev}}, \bibinfo {author} {\bibfnamefont {E.}~\bibnamefont {Schwegler}},
  \bibinfo {author} {\bibfnamefont {T.}~\bibnamefont {Ogitsu}}, \ and\ \bibinfo
  {author} {\bibfnamefont {G.}~\bibnamefont {Galli}},\ }\href {\doibase
  10.1038/nature02968} {\bibfield  {journal} {\bibinfo  {journal} {Nature}\
  }\textbf {\bibinfo {volume} {431}},\ \bibinfo {pages} {669} (\bibinfo {year}
  {2004})}\BibitemShut {NoStop}%
\bibitem [{\citenamefont {Pickard}\ and\ \citenamefont
  {Needs}(2007)}]{Pickard2007}%
  \BibitemOpen
  \bibfield  {author} {\bibinfo {author} {\bibfnamefont {C.~J.}\ \bibnamefont
  {Pickard}}\ and\ \bibinfo {author} {\bibfnamefont {R.~J.}\ \bibnamefont
  {Needs}},\ }\href {\doibase 10.1038/nphys625} {\bibfield  {journal} {\bibinfo
   {journal} {Nat.Phys.}\ }\textbf {\bibinfo {volume} {3}},\ \bibinfo {pages}
  {473} (\bibinfo {year} {2007})}\BibitemShut {NoStop}%
\bibitem [{\citenamefont {McMahon}\ and\ \citenamefont
  {Ceperley}(2011{\natexlab{a}})}]{mcmahon2011ground}%
  \BibitemOpen
  \bibfield  {author} {\bibinfo {author} {\bibfnamefont {J.~M.}\ \bibnamefont
  {McMahon}}\ and\ \bibinfo {author} {\bibfnamefont {D.~M.}\ \bibnamefont
  {Ceperley}},\ }\href {\doibase 10.1103/PhysRevLett.106.165302} {\bibfield
  {journal} {\bibinfo  {journal} {Phys.Rev.Lett.}\ }\textbf {\bibinfo {volume}
  {106}},\ \bibinfo {pages} {165302} (\bibinfo {year}
  {2011}{\natexlab{a}})}\BibitemShut {NoStop}%
\bibitem [{\citenamefont {Zha}\ \emph {et~al.}(2012)\citenamefont {Zha},
  \citenamefont {Liu},\ and\ \citenamefont {Hemley}}]{zha2012synchrotron}%
  \BibitemOpen
  \bibfield  {author} {\bibinfo {author} {\bibfnamefont {C.-S.}\ \bibnamefont
  {Zha}}, \bibinfo {author} {\bibfnamefont {Z.}~\bibnamefont {Liu}}, \ and\
  \bibinfo {author} {\bibfnamefont {R.~J.}\ \bibnamefont {Hemley}},\ }\href
  {\doibase 10.1103/PhysRevLett.108.146402} {\bibfield  {journal} {\bibinfo
  {journal} {Phys.Rev.Lett.}\ }\textbf {\bibinfo {volume} {108}},\ \bibinfo
  {pages} {146402} (\bibinfo {year} {2012})}\BibitemShut {NoStop}%
\bibitem [{\citenamefont {Chen}\ \emph {et~al.}(2013)\citenamefont {Chen},
  \citenamefont {Li}, \citenamefont {Zhang}, \citenamefont {Probert},
  \citenamefont {Pickard}, \citenamefont {Needs}, \citenamefont {Michaelides},\
  and\ \citenamefont {Wang}}]{chen2013quantum}%
  \BibitemOpen
  \bibfield  {author} {\bibinfo {author} {\bibfnamefont {J.}~\bibnamefont
  {Chen}}, \bibinfo {author} {\bibfnamefont {X.-Z.}\ \bibnamefont {Li}},
  \bibinfo {author} {\bibfnamefont {Q.}~\bibnamefont {Zhang}}, \bibinfo
  {author} {\bibfnamefont {M.~I.~J.}\ \bibnamefont {Probert}}, \bibinfo
  {author} {\bibfnamefont {C.~J.}\ \bibnamefont {Pickard}}, \bibinfo {author}
  {\bibfnamefont {R.~J.}\ \bibnamefont {Needs}}, \bibinfo {author}
  {\bibfnamefont {A.}~\bibnamefont {Michaelides}}, \ and\ \bibinfo {author}
  {\bibfnamefont {E.}~\bibnamefont {Wang}},\ }\href {\doibase
  10.1038/ncomms3064} {\bibfield  {journal} {\bibinfo  {journal} {Nat.
  Commun.}\ }\textbf {\bibinfo {volume} {4}},\ \bibinfo {pages} {2064}
  (\bibinfo {year} {2013})}\BibitemShut {NoStop}%
\bibitem [{\citenamefont {Azadi}\ \emph {et~al.}(2014)\citenamefont {Azadi},
  \citenamefont {Monserrat}, \citenamefont {Foulkes},\ and\ \citenamefont
  {Needs}}]{azadi2014dissociation}%
  \BibitemOpen
  \bibfield  {author} {\bibinfo {author} {\bibfnamefont {S.}~\bibnamefont
  {Azadi}}, \bibinfo {author} {\bibfnamefont {B.}~\bibnamefont {Monserrat}},
  \bibinfo {author} {\bibfnamefont {W.~M.~C.}\ \bibnamefont {Foulkes}}, \ and\
  \bibinfo {author} {\bibfnamefont {R.~J.}\ \bibnamefont {Needs}},\ }\href
  {\doibase 10.1103/PhysRevLett.112.165501} {\bibfield  {journal} {\bibinfo
  {journal} {Phys.Rev.Lett.}\ }\textbf {\bibinfo {volume} {112}},\ \bibinfo
  {pages} {165501} (\bibinfo {year} {2014})}\BibitemShut {NoStop}%
\bibitem [{\citenamefont {Dias}\ and\ \citenamefont
  {Silvera}(2017)}]{dias2017observation}%
  \BibitemOpen
  \bibfield  {author} {\bibinfo {author} {\bibfnamefont {R.~P.}\ \bibnamefont
  {Dias}}\ and\ \bibinfo {author} {\bibfnamefont {I.~F.}\ \bibnamefont
  {Silvera}},\ }\href {\doibase 10.1126/science.aal1579} {\bibfield  {journal}
  {\bibinfo  {journal} {Science}\ }\textbf {\bibinfo {volume} {355}},\ \bibinfo
  {pages} {715} (\bibinfo {year} {2017})}\BibitemShut {NoStop}%
\bibitem [{\citenamefont {McMinis}\ \emph {et~al.}(2015)\citenamefont
  {McMinis}, \citenamefont {Clay~III}, \citenamefont {Lee},\ and\ \citenamefont
  {Morales}}]{McMinis2015}%
  \BibitemOpen
  \bibfield  {author} {\bibinfo {author} {\bibfnamefont {J.}~\bibnamefont
  {McMinis}}, \bibinfo {author} {\bibfnamefont {R.~C.}\ \bibnamefont
  {Clay~III}}, \bibinfo {author} {\bibfnamefont {D.}~\bibnamefont {Lee}}, \
  and\ \bibinfo {author} {\bibfnamefont {M.~A.}\ \bibnamefont {Morales}},\
  }\href {\doibase 10.1103/PhysRevLett.114.105305} {\bibfield  {journal}
  {\bibinfo  {journal} {Phys.Rev.Lett.}\ }\textbf {\bibinfo {volume} {114}},\
  \bibinfo {pages} {105305} (\bibinfo {year} {2015})}\BibitemShut {NoStop}%
\bibitem [{\citenamefont {Cudazzo}\ \emph {et~al.}(2008)\citenamefont
  {Cudazzo}, \citenamefont {Profeta}, \citenamefont {Sanna}, \citenamefont
  {Floris}, \citenamefont {Continenza}, \citenamefont {Massidda},\ and\
  \citenamefont {Gross}}]{cudazzo2008ab}%
  \BibitemOpen
  \bibfield  {author} {\bibinfo {author} {\bibfnamefont {P.}~\bibnamefont
  {Cudazzo}}, \bibinfo {author} {\bibfnamefont {G.}~\bibnamefont {Profeta}},
  \bibinfo {author} {\bibfnamefont {A.}~\bibnamefont {Sanna}}, \bibinfo
  {author} {\bibfnamefont {A.}~\bibnamefont {Floris}}, \bibinfo {author}
  {\bibfnamefont {A.}~\bibnamefont {Continenza}}, \bibinfo {author}
  {\bibfnamefont {S.}~\bibnamefont {Massidda}}, \ and\ \bibinfo {author}
  {\bibfnamefont {E.~K.~U.}\ \bibnamefont {Gross}},\ }\href {\doibase
  10.1103/PhysRevLett.100.257001} {\bibfield  {journal} {\bibinfo  {journal}
  {Phys.Rev.Lett.}\ }\textbf {\bibinfo {volume} {100}},\ \bibinfo {pages}
  {257001} (\bibinfo {year} {2008})}\BibitemShut {NoStop}%
\bibitem [{\citenamefont {Borinaga}\ \emph {et~al.}(2016)\citenamefont
  {Borinaga}, \citenamefont {Errea}, \citenamefont {Calandra}, \citenamefont
  {Mauri},\ and\ \citenamefont {Bergara}}]{borinaga2016anharmonic}%
  \BibitemOpen
  \bibfield  {author} {\bibinfo {author} {\bibfnamefont {M.}~\bibnamefont
  {Borinaga}}, \bibinfo {author} {\bibfnamefont {I.}~\bibnamefont {Errea}},
  \bibinfo {author} {\bibfnamefont {M.}~\bibnamefont {Calandra}}, \bibinfo
  {author} {\bibfnamefont {F.}~\bibnamefont {Mauri}}, \ and\ \bibinfo {author}
  {\bibfnamefont {A.}~\bibnamefont {Bergara}},\ }\href {\doibase
  10.1103/PhysRevB.93.174308} {\bibfield  {journal} {\bibinfo  {journal}
  {Phys.Rev.B}\ }\textbf {\bibinfo {volume} {93}},\ \bibinfo {pages} {174308}
  (\bibinfo {year} {2016})}\BibitemShut {NoStop}%
\bibitem [{\citenamefont {McMahon}\ and\ \citenamefont
  {Ceperley}(2011{\natexlab{b}})}]{mcmahon2011high}%
  \BibitemOpen
  \bibfield  {author} {\bibinfo {author} {\bibfnamefont {J.~M.}\ \bibnamefont
  {McMahon}}\ and\ \bibinfo {author} {\bibfnamefont {D.~M.}\ \bibnamefont
  {Ceperley}},\ }\href {\doibase 10.1103/PhysRevB.84.144515} {\bibfield
  {journal} {\bibinfo  {journal} {Phys.Rev.B}\ }\textbf {\bibinfo {volume}
  {84}},\ \bibinfo {pages} {144515} (\bibinfo {year}
  {2011}{\natexlab{b}})}\BibitemShut {NoStop}%
\bibitem [{\citenamefont {Babaev}\ \emph {et~al.}(2004)\citenamefont {Babaev},
  \citenamefont {Sudb{\o}},\ and\ \citenamefont
  {Ashcroft}}]{babaev2004superconductor}%
  \BibitemOpen
  \bibfield  {author} {\bibinfo {author} {\bibfnamefont {E.}~\bibnamefont
  {Babaev}}, \bibinfo {author} {\bibfnamefont {A.}~\bibnamefont {Sudb{\o}}}, \
  and\ \bibinfo {author} {\bibfnamefont {N.~W.}\ \bibnamefont {Ashcroft}},\
  }\href {\doibase 10.1038/nature02910} {\bibfield  {journal} {\bibinfo
  {journal} {Nature}\ }\textbf {\bibinfo {volume} {431}},\ \bibinfo {pages}
  {666} (\bibinfo {year} {2004})}\BibitemShut {NoStop}%
\bibitem [{\citenamefont {Silvera}\ and\ \citenamefont
  {Cole}(2010)}]{silvera2010metallic}%
  \BibitemOpen
  \bibfield  {author} {\bibinfo {author} {\bibfnamefont {I.~F.}\ \bibnamefont
  {Silvera}}\ and\ \bibinfo {author} {\bibfnamefont {J.~W.}\ \bibnamefont
  {Cole}},\ }in\ \href
  {http://iopscience.iop.org/article/10.1088/1742-6596/215/1/012194/meta}
  {\emph {\bibinfo {booktitle} {J. Phys. Conf. Ser.}}},\ Vol.\ \bibinfo
  {volume} {215}\ (\bibinfo {organization} {IOP Publishing},\ \bibinfo {year}
  {2010})\ p.\ \bibinfo {pages} {012194}\BibitemShut {NoStop}%
\bibitem [{\citenamefont {McMahon}\ \emph {et~al.}(2012)\citenamefont
  {McMahon}, \citenamefont {Morales}, \citenamefont {Pierleoni},\ and\
  \citenamefont {Ceperley}}]{Ceperley2012}%
  \BibitemOpen
  \bibfield  {author} {\bibinfo {author} {\bibfnamefont {J.~M.}\ \bibnamefont
  {McMahon}}, \bibinfo {author} {\bibfnamefont {M.~A.}\ \bibnamefont
  {Morales}}, \bibinfo {author} {\bibfnamefont {C.}~\bibnamefont {Pierleoni}},
  \ and\ \bibinfo {author} {\bibfnamefont {D.~M.}\ \bibnamefont {Ceperley}},\
  }\href {\doibase 10.1103/RevModPhys.84.1607} {\bibfield  {journal} {\bibinfo
  {journal} {Rev. Mod. Phys.}\ }\textbf {\bibinfo {volume} {84}},\ \bibinfo
  {pages} {1607} (\bibinfo {year} {2012})}\BibitemShut {NoStop}%
\bibitem [{\citenamefont {Mao}\ \emph {et~al.}(1988)\citenamefont {Mao},
  \citenamefont {Jephcoat}, \citenamefont {Hemley}, \citenamefont {Finger},
  \citenamefont {Zha}, \citenamefont {Hazen},\ and\ \citenamefont
  {Cox}}]{mao1988synchrotron}%
  \BibitemOpen
  \bibfield  {author} {\bibinfo {author} {\bibfnamefont {H.~K.}\ \bibnamefont
  {Mao}}, \bibinfo {author} {\bibfnamefont {A.~P.}\ \bibnamefont {Jephcoat}},
  \bibinfo {author} {\bibfnamefont {R.~J.}\ \bibnamefont {Hemley}}, \bibinfo
  {author} {\bibfnamefont {L.~W.}\ \bibnamefont {Finger}}, \bibinfo {author}
  {\bibfnamefont {C.~S.}\ \bibnamefont {Zha}}, \bibinfo {author} {\bibfnamefont
  {R.~M.}\ \bibnamefont {Hazen}}, \ and\ \bibinfo {author} {\bibfnamefont
  {D.~E.}\ \bibnamefont {Cox}},\ }\href {\doibase
  10.1126/science.239.4844.1131} {\bibfield  {journal} {\bibinfo  {journal}
  {Science}\ }\textbf {\bibinfo {volume} {239}},\ \bibinfo {pages} {1131}
  (\bibinfo {year} {1988})}\BibitemShut {NoStop}%
\bibitem [{\citenamefont {Akahama}\ \emph {et~al.}(2010)\citenamefont
  {Akahama}, \citenamefont {Nishimura}, \citenamefont {Kawamura}, \citenamefont
  {Hirao}, \citenamefont {Ohishi},\ and\ \citenamefont
  {Takemura}}]{akahama2010evidence}%
  \BibitemOpen
  \bibfield  {author} {\bibinfo {author} {\bibfnamefont {Y.}~\bibnamefont
  {Akahama}}, \bibinfo {author} {\bibfnamefont {M.}~\bibnamefont {Nishimura}},
  \bibinfo {author} {\bibfnamefont {H.}~\bibnamefont {Kawamura}}, \bibinfo
  {author} {\bibfnamefont {N.}~\bibnamefont {Hirao}}, \bibinfo {author}
  {\bibfnamefont {Y.}~\bibnamefont {Ohishi}}, \ and\ \bibinfo {author}
  {\bibfnamefont {K.}~\bibnamefont {Takemura}},\ }\href {\doibase
  10.1103/PhysRevB.82.060101} {\bibfield  {journal} {\bibinfo  {journal}
  {Phys.Rev.B}\ }\textbf {\bibinfo {volume} {82}},\ \bibinfo {pages} {060101}
  (\bibinfo {year} {2010})}\BibitemShut {NoStop}%
\bibitem [{\citenamefont {Lorenzana}\ \emph {et~al.}(1989)\citenamefont
  {Lorenzana}, \citenamefont {Silvera},\ and\ \citenamefont
  {Goettel}}]{lorenzana1989evidence}%
  \BibitemOpen
  \bibfield  {author} {\bibinfo {author} {\bibfnamefont {H.~E.}\ \bibnamefont
  {Lorenzana}}, \bibinfo {author} {\bibfnamefont {I.~F.}\ \bibnamefont
  {Silvera}}, \ and\ \bibinfo {author} {\bibfnamefont {K.~A.}\ \bibnamefont
  {Goettel}},\ }\href {\doibase 10.1103/PhysRevLett.63.2080} {\bibfield
  {journal} {\bibinfo  {journal} {Phys.Rev.Lett.}\ }\textbf {\bibinfo {volume}
  {63}},\ \bibinfo {pages} {2080} (\bibinfo {year} {1989})}\BibitemShut
  {NoStop}%
\bibitem [{\citenamefont {Hanfland}\ \emph {et~al.}(1993)\citenamefont
  {Hanfland}, \citenamefont {Hemley},\ and\ \citenamefont
  {Mao}}]{hanfland1993novel}%
  \BibitemOpen
  \bibfield  {author} {\bibinfo {author} {\bibfnamefont {M.}~\bibnamefont
  {Hanfland}}, \bibinfo {author} {\bibfnamefont {R.~J.}\ \bibnamefont
  {Hemley}}, \ and\ \bibinfo {author} {\bibfnamefont {H.~K.}\ \bibnamefont
  {Mao}},\ }\href {\doibase 10.1103/PhysRevLett.108.146402} {\bibfield
  {journal} {\bibinfo  {journal} {Phys.Rev.Lett.}\ }\textbf {\bibinfo {volume}
  {70}},\ \bibinfo {pages} {3760} (\bibinfo {year} {1993})}\BibitemShut
  {NoStop}%
\bibitem [{\citenamefont {Lorenzana}\ \emph {et~al.}(1990)\citenamefont
  {Lorenzana}, \citenamefont {Silvera},\ and\ \citenamefont
  {Goettel}}]{lorenzana1990orientational}%
  \BibitemOpen
  \bibfield  {author} {\bibinfo {author} {\bibfnamefont {H.~E.}\ \bibnamefont
  {Lorenzana}}, \bibinfo {author} {\bibfnamefont {I.~F.}\ \bibnamefont
  {Silvera}}, \ and\ \bibinfo {author} {\bibfnamefont {K.~A.}\ \bibnamefont
  {Goettel}},\ }\href {\doibase 10.1103/PhysRevLett.64.1939} {\bibfield
  {journal} {\bibinfo  {journal} {Phys.Rev.Lett.}\ }\textbf {\bibinfo {volume}
  {64}},\ \bibinfo {pages} {1939} (\bibinfo {year} {1990})}\BibitemShut
  {NoStop}%
\bibitem [{\citenamefont {Hanfland}\ \emph {et~al.}(1992)\citenamefont
  {Hanfland}, \citenamefont {Hemley}, \citenamefont {Mao},\ and\ \citenamefont
  {Williams}}]{hanfland1992synchrotron}%
  \BibitemOpen
  \bibfield  {author} {\bibinfo {author} {\bibfnamefont {M.}~\bibnamefont
  {Hanfland}}, \bibinfo {author} {\bibfnamefont {R.~J.}\ \bibnamefont
  {Hemley}}, \bibinfo {author} {\bibfnamefont {H.~K.}\ \bibnamefont {Mao}}, \
  and\ \bibinfo {author} {\bibfnamefont {G.~P.}\ \bibnamefont {Williams}},\
  }\href {\doibase 10.1103/PhysRevLett.69.1129} {\bibfield  {journal} {\bibinfo
   {journal} {Phys.Rev.Lett.}\ }\textbf {\bibinfo {volume} {69}},\ \bibinfo
  {pages} {1129} (\bibinfo {year} {1992})}\BibitemShut {NoStop}%
\bibitem [{\citenamefont {Hemley}\ \emph {et~al.}(1990)\citenamefont {Hemley},
  \citenamefont {Mao},\ and\ \citenamefont {Shu}}]{hemley1990low}%
  \BibitemOpen
  \bibfield  {author} {\bibinfo {author} {\bibfnamefont {R.~J.}\ \bibnamefont
  {Hemley}}, \bibinfo {author} {\bibfnamefont {H.~K.}\ \bibnamefont {Mao}}, \
  and\ \bibinfo {author} {\bibfnamefont {J.~F.}\ \bibnamefont {Shu}},\ }\href
  {\doibase 10.1103/PhysRevLett.65.2670} {\bibfield  {journal} {\bibinfo
  {journal} {Phys.Rev.Lett.}\ }\textbf {\bibinfo {volume} {65}},\ \bibinfo
  {pages} {2670} (\bibinfo {year} {1990})}\BibitemShut {NoStop}%
\bibitem [{\citenamefont {Goncharov}\ \emph {et~al.}(1996)\citenamefont
  {Goncharov}, \citenamefont {Eggert}, \citenamefont {Mazin}, \citenamefont
  {Hemley},\ and\ \citenamefont {Mao}}]{goncharov1996raman}%
  \BibitemOpen
  \bibfield  {author} {\bibinfo {author} {\bibfnamefont {A.~F.}\ \bibnamefont
  {Goncharov}}, \bibinfo {author} {\bibfnamefont {J.~H.}\ \bibnamefont
  {Eggert}}, \bibinfo {author} {\bibfnamefont {I.~I.}\ \bibnamefont {Mazin}},
  \bibinfo {author} {\bibfnamefont {R.~J.}\ \bibnamefont {Hemley}}, \ and\
  \bibinfo {author} {\bibfnamefont {H.~K.}\ \bibnamefont {Mao}},\ }\href
  {\doibase 10.1103/PhysRevB.54.R15590} {\bibfield  {journal} {\bibinfo
  {journal} {Phys.Rev.B}\ }\textbf {\bibinfo {volume} {54}},\ \bibinfo {pages}
  {R15590} (\bibinfo {year} {1996})}\BibitemShut {NoStop}%
\bibitem [{\citenamefont {Hemley}\ \emph {et~al.}(1997)\citenamefont {Hemley},
  \citenamefont {Mazin}, \citenamefont {Goncharov},\ and\ \citenamefont
  {Mao}}]{hemley1997vibron}%
  \BibitemOpen
  \bibfield  {author} {\bibinfo {author} {\bibfnamefont {R.~J.}\ \bibnamefont
  {Hemley}}, \bibinfo {author} {\bibfnamefont {I.~I.}\ \bibnamefont {Mazin}},
  \bibinfo {author} {\bibfnamefont {A.~F.}\ \bibnamefont {Goncharov}}, \ and\
  \bibinfo {author} {\bibfnamefont {H.~K.}\ \bibnamefont {Mao}},\ }\href
  {\doibase 10.1209/epl/i1997-00163-6} {\bibfield  {journal} {\bibinfo
  {journal} {EPL}\ }\textbf {\bibinfo {volume} {37}},\ \bibinfo {pages} {403}
  (\bibinfo {year} {1997})}\BibitemShut {NoStop}%
\bibitem [{\citenamefont {Gregoryanz}\ \emph {et~al.}(2003)\citenamefont
  {Gregoryanz}, \citenamefont {Goncharov}, \citenamefont {Matsuishi},
  \citenamefont {Mao},\ and\ \citenamefont {Hemley}}]{gregoryanz2003raman}%
  \BibitemOpen
  \bibfield  {author} {\bibinfo {author} {\bibfnamefont {E.}~\bibnamefont
  {Gregoryanz}}, \bibinfo {author} {\bibfnamefont {A.~F.}\ \bibnamefont
  {Goncharov}}, \bibinfo {author} {\bibfnamefont {K.}~\bibnamefont
  {Matsuishi}}, \bibinfo {author} {\bibfnamefont {H.~K.}\ \bibnamefont {Mao}},
  \ and\ \bibinfo {author} {\bibfnamefont {R.~J.}\ \bibnamefont {Hemley}},\
  }\href {\doibase 10.1103/PhysRevLett.90.175701} {\bibfield  {journal}
  {\bibinfo  {journal} {Phys.Rev.Lett.}\ }\textbf {\bibinfo {volume} {90}},\
  \bibinfo {pages} {175701} (\bibinfo {year} {2003})}\BibitemShut {NoStop}%
\bibitem [{\citenamefont {Zha}\ \emph {et~al.}(2013)\citenamefont {Zha},
  \citenamefont {Liu}, \citenamefont {Ahart}, \citenamefont {Boehler},\ and\
  \citenamefont {Hemley}}]{zha2013high}%
  \BibitemOpen
  \bibfield  {author} {\bibinfo {author} {\bibfnamefont {C.-s.}\ \bibnamefont
  {Zha}}, \bibinfo {author} {\bibfnamefont {Z.}~\bibnamefont {Liu}}, \bibinfo
  {author} {\bibfnamefont {M.}~\bibnamefont {Ahart}}, \bibinfo {author}
  {\bibfnamefont {R.}~\bibnamefont {Boehler}}, \ and\ \bibinfo {author}
  {\bibfnamefont {R.~J.}\ \bibnamefont {Hemley}},\ }\href {\doibase
  10.1103/PhysRevLett.110.217402} {\bibfield  {journal} {\bibinfo  {journal}
  {Phys.Rev.Lett.}\ }\textbf {\bibinfo {volume} {110}},\ \bibinfo {pages}
  {217402} (\bibinfo {year} {2013})}\BibitemShut {NoStop}%
\bibitem [{\citenamefont {Dalladay-Simpson}\ \emph {et~al.}(2016)\citenamefont
  {Dalladay-Simpson}, \citenamefont {Howie},\ and\ \citenamefont
  {Gregoryanz}}]{dalladay2016evidence}%
  \BibitemOpen
  \bibfield  {author} {\bibinfo {author} {\bibfnamefont {P.}~\bibnamefont
  {Dalladay-Simpson}}, \bibinfo {author} {\bibfnamefont {R.~T.}\ \bibnamefont
  {Howie}}, \ and\ \bibinfo {author} {\bibfnamefont {E.}~\bibnamefont
  {Gregoryanz}},\ }\href {\doibase 10.1038/nature16164} {\bibfield  {journal}
  {\bibinfo  {journal} {Nature}\ }\textbf {\bibinfo {volume} {529}},\ \bibinfo
  {pages} {63} (\bibinfo {year} {2016})}\BibitemShut {NoStop}%
\bibitem [{\citenamefont {Howie}\ \emph {et~al.}(2015)\citenamefont {Howie},
  \citenamefont {Dalladay-Simpson},\ and\ \citenamefont
  {Gregoryanz}}]{howie2015raman}%
  \BibitemOpen
  \bibfield  {author} {\bibinfo {author} {\bibfnamefont {R.~T.}\ \bibnamefont
  {Howie}}, \bibinfo {author} {\bibfnamefont {P.}~\bibnamefont
  {Dalladay-Simpson}}, \ and\ \bibinfo {author} {\bibfnamefont
  {E.}~\bibnamefont {Gregoryanz}},\ }\href {\doibase 10.1038/nmat4213}
  {\bibfield  {journal} {\bibinfo  {journal} {Nat.Mater.}\ }\textbf {\bibinfo
  {volume} {14}},\ \bibinfo {pages} {495} (\bibinfo {year} {2015})}\BibitemShut
  {NoStop}%
\bibitem [{\citenamefont {Zha}\ \emph {et~al.}(2014)\citenamefont {Zha},
  \citenamefont {Cohen}, \citenamefont {Mao},\ and\ \citenamefont
  {Hemley}}]{zha2014raman}%
  \BibitemOpen
  \bibfield  {author} {\bibinfo {author} {\bibfnamefont {C.-s.}\ \bibnamefont
  {Zha}}, \bibinfo {author} {\bibfnamefont {R.~E.}\ \bibnamefont {Cohen}},
  \bibinfo {author} {\bibfnamefont {H.~K.}\ \bibnamefont {Mao}}, \ and\
  \bibinfo {author} {\bibfnamefont {R.~J.}\ \bibnamefont {Hemley}},\ }\href
  {\doibase 10.1073/pnas.1402737111} {\bibfield  {journal} {\bibinfo  {journal}
  {‎Proc. Natl. Acad. Sci. U.S.A}\ }\textbf {\bibinfo {volume} {111}},\
  \bibinfo {pages} {4792} (\bibinfo {year} {2014})}\BibitemShut {NoStop}%
\bibitem [{\citenamefont {Drozdov}\ \emph {et~al.}(2015)\citenamefont
  {Drozdov}, \citenamefont {Eremets}, \citenamefont {Troyan}, \citenamefont
  {Ksenofontov},\ and\ \citenamefont {Shylin}}]{drozdov2015conventional}%
  \BibitemOpen
  \bibfield  {author} {\bibinfo {author} {\bibfnamefont {A.~P.}\ \bibnamefont
  {Drozdov}}, \bibinfo {author} {\bibfnamefont {M.~I.}\ \bibnamefont
  {Eremets}}, \bibinfo {author} {\bibfnamefont {I.~A.}\ \bibnamefont {Troyan}},
  \bibinfo {author} {\bibfnamefont {V.}~\bibnamefont {Ksenofontov}}, \ and\
  \bibinfo {author} {\bibfnamefont {S.~I.}\ \bibnamefont {Shylin}},\ }\href
  {\doibase 10.1038/nature14964} {\bibfield  {journal} {\bibinfo  {journal}
  {Nature}\ }\textbf {\bibinfo {volume} {525}},\ \bibinfo {pages} {73}
  (\bibinfo {year} {2015})}\BibitemShut {NoStop}%
\bibitem [{\citenamefont {Eremets}\ and\ \citenamefont
  {Troyan}(2011)}]{eremets2011conductive}%
  \BibitemOpen
  \bibfield  {author} {\bibinfo {author} {\bibfnamefont {M.~I.}\ \bibnamefont
  {Eremets}}\ and\ \bibinfo {author} {\bibfnamefont {I.~A.}\ \bibnamefont
  {Troyan}},\ }\href {\doibase 10.1038/nmat3175} {\bibfield  {journal}
  {\bibinfo  {journal} {Nat.Mater.}\ }\textbf {\bibinfo {volume} {10}},\
  \bibinfo {pages} {927} (\bibinfo {year} {2011})}\BibitemShut {NoStop}%
\bibitem [{\citenamefont {Eremets}\ \emph {et~al.}(2017)\citenamefont
  {Eremets}, \citenamefont {Drozdov}, \citenamefont {Kong},\ and\ \citenamefont
  {Wang}}]{eremets2017molecular}%
  \BibitemOpen
  \bibfield  {author} {\bibinfo {author} {\bibfnamefont {M.~I.}\ \bibnamefont
  {Eremets}}, \bibinfo {author} {\bibfnamefont {A.~P.}\ \bibnamefont
  {Drozdov}}, \bibinfo {author} {\bibfnamefont {P.~P.}\ \bibnamefont {Kong}}, \
  and\ \bibinfo {author} {\bibfnamefont {H.}~\bibnamefont {Wang}},\ }\href@noop
  {} {\bibfield  {journal} {\bibinfo  {journal} {arXiv:1708.05217}\ } (\bibinfo
  {year} {2017})}\BibitemShut {NoStop}%
\bibitem [{\citenamefont {Mao}\ \emph {et~al.}(1990)\citenamefont {Mao},
  \citenamefont {Hemley},\ and\ \citenamefont {Hanfland}}]{mao1990infrared}%
  \BibitemOpen
  \bibfield  {author} {\bibinfo {author} {\bibfnamefont {H.~K.}\ \bibnamefont
  {Mao}}, \bibinfo {author} {\bibfnamefont {R.~J.}\ \bibnamefont {Hemley}}, \
  and\ \bibinfo {author} {\bibfnamefont {M.}~\bibnamefont {Hanfland}},\ }\href
  {\doibase 10.1103/PhysRevLett.65.484} {\bibfield  {journal} {\bibinfo
  {journal} {Phys.Rev.Lett.}\ }\textbf {\bibinfo {volume} {65}},\ \bibinfo
  {pages} {484} (\bibinfo {year} {1990})}\BibitemShut {NoStop}%
\bibitem [{\citenamefont {Eggert}\ \emph {et~al.}(1991)\citenamefont {Eggert},
  \citenamefont {Moshary}, \citenamefont {Evans}, \citenamefont {Lorenzana},
  \citenamefont {Goettel}, \citenamefont {Silvera},\ and\ \citenamefont
  {Moss}}]{eggert1991absorption}%
  \BibitemOpen
  \bibfield  {author} {\bibinfo {author} {\bibfnamefont {J.~H.}\ \bibnamefont
  {Eggert}}, \bibinfo {author} {\bibfnamefont {F.}~\bibnamefont {Moshary}},
  \bibinfo {author} {\bibfnamefont {W.~J.}\ \bibnamefont {Evans}}, \bibinfo
  {author} {\bibfnamefont {H.~E.}\ \bibnamefont {Lorenzana}}, \bibinfo {author}
  {\bibfnamefont {K.~A.}\ \bibnamefont {Goettel}}, \bibinfo {author}
  {\bibfnamefont {I.~F.}\ \bibnamefont {Silvera}}, \ and\ \bibinfo {author}
  {\bibfnamefont {W.~C.}\ \bibnamefont {Moss}},\ }\href {\doibase
  10.1103/PhysRevLett.66.193} {\bibfield  {journal} {\bibinfo  {journal}
  {Phys.Rev.Lett.}\ }\textbf {\bibinfo {volume} {66}},\ \bibinfo {pages} {193}
  (\bibinfo {year} {1991})}\BibitemShut {NoStop}%
\bibitem [{\citenamefont {Hemley}\ \emph {et~al.}(1991)\citenamefont {Hemley},
  \citenamefont {Hanfland},\ and\ \citenamefont {Mao}}]{hemley1991high}%
  \BibitemOpen
  \bibfield  {author} {\bibinfo {author} {\bibfnamefont {R.~J.}\ \bibnamefont
  {Hemley}}, \bibinfo {author} {\bibfnamefont {M.}~\bibnamefont {Hanfland}}, \
  and\ \bibinfo {author} {\bibfnamefont {H.~K.}\ \bibnamefont {Mao}},\ }\href
  {\doibase 10.1038/350488a0} {\bibfield  {journal} {\bibinfo  {journal}
  {Nature}\ }\textbf {\bibinfo {volume} {350}},\ \bibinfo {pages} {488}
  (\bibinfo {year} {1991})}\BibitemShut {NoStop}%
\bibitem [{\citenamefont {Mao}\ and\ \citenamefont
  {Hemley}(1989)}]{mao1989optical}%
  \BibitemOpen
  \bibfield  {author} {\bibinfo {author} {\bibfnamefont {H.~K.}\ \bibnamefont
  {Mao}}\ and\ \bibinfo {author} {\bibfnamefont {R.~J.}\ \bibnamefont
  {Hemley}},\ }\href {\doibase 10.1126/science.244.4911.1462} {\bibfield
  {journal} {\bibinfo  {journal} {Science}\ }\textbf {\bibinfo {volume}
  {244}},\ \bibinfo {pages} {1462} (\bibinfo {year} {1989})}\BibitemShut
  {NoStop}%
\bibitem [{\citenamefont {Howie}\ \emph {et~al.}(2012)\citenamefont {Howie},
  \citenamefont {Guillaume}, \citenamefont {Scheler}, \citenamefont
  {Goncharov},\ and\ \citenamefont {Gregoryanz}}]{howie2012mixed}%
  \BibitemOpen
  \bibfield  {author} {\bibinfo {author} {\bibfnamefont {R.~T.}\ \bibnamefont
  {Howie}}, \bibinfo {author} {\bibfnamefont {C.~L.}\ \bibnamefont
  {Guillaume}}, \bibinfo {author} {\bibfnamefont {T.}~\bibnamefont {Scheler}},
  \bibinfo {author} {\bibfnamefont {A.~F.}\ \bibnamefont {Goncharov}}, \ and\
  \bibinfo {author} {\bibfnamefont {E.}~\bibnamefont {Gregoryanz}},\ }\href
  {\doibase 10.1103/PhysRevLett.108.125501} {\bibfield  {journal} {\bibinfo
  {journal} {Phys.Rev.Lett.}\ }\textbf {\bibinfo {volume} {108}},\ \bibinfo
  {pages} {125501} (\bibinfo {year} {2012})}\BibitemShut {NoStop}%
\bibitem [{\citenamefont {Loubeyre}\ \emph {et~al.}(2002)\citenamefont
  {Loubeyre}, \citenamefont {Occelli},\ and\ \citenamefont
  {LeToullec}}]{loubeyre2002optical}%
  \BibitemOpen
  \bibfield  {author} {\bibinfo {author} {\bibfnamefont {P.}~\bibnamefont
  {Loubeyre}}, \bibinfo {author} {\bibfnamefont {F.}~\bibnamefont {Occelli}}, \
  and\ \bibinfo {author} {\bibfnamefont {R.}~\bibnamefont {LeToullec}},\ }\href
  {\doibase Optical studies of solid hydrogen to 320 GPa and evidence for
  black hydrogen} {\bibfield  {journal} {\bibinfo  {journal} {Nature}\ }\textbf
  {\bibinfo {volume} {416}},\ \bibinfo {pages} {613} (\bibinfo {year}
  {2002})}\BibitemShut {NoStop}%
\bibitem [{\citenamefont {Evans}\ and\ \citenamefont
  {Silvera}(1998)}]{evans1998index}%
  \BibitemOpen
  \bibfield  {author} {\bibinfo {author} {\bibfnamefont {W.~J.}\ \bibnamefont
  {Evans}}\ and\ \bibinfo {author} {\bibfnamefont {I.~F.}\ \bibnamefont
  {Silvera}},\ }\href {\doibase 10.1103/PhysRevB.57.14105} {\bibfield
  {journal} {\bibinfo  {journal} {Phys.Rev.B}\ }\textbf {\bibinfo {volume}
  {57}},\ \bibinfo {pages} {14105} (\bibinfo {year} {1998})}\BibitemShut
  {NoStop}%
\bibitem [{\citenamefont {Oganov}\ and\ \citenamefont
  {Glass}(2006)}]{oganov2006crystal}%
  \BibitemOpen
  \bibfield  {author} {\bibinfo {author} {\bibfnamefont {A.~R.}\ \bibnamefont
  {Oganov}}\ and\ \bibinfo {author} {\bibfnamefont {C.~W.}\ \bibnamefont
  {Glass}},\ }\href {\doibase 10.1063/1.2210932} {\bibfield  {journal}
  {\bibinfo  {journal} {J.Chem.Phys.}\ }\textbf {\bibinfo {volume} {124}},\
  \bibinfo {pages} {244704} (\bibinfo {year} {2006})}\BibitemShut {NoStop}%
\bibitem [{\citenamefont {Pickard}\ and\ \citenamefont
  {Needs}(2006)}]{pickard2006high}%
  \BibitemOpen
  \bibfield  {author} {\bibinfo {author} {\bibfnamefont {C.~J.}\ \bibnamefont
  {Pickard}}\ and\ \bibinfo {author} {\bibfnamefont {R.~J.}\ \bibnamefont
  {Needs}},\ }\href {\doibase 10.1103/PhysRevLett.97.045504} {\bibfield
  {journal} {\bibinfo  {journal} {Phys.Rev.Lett.}\ }\textbf {\bibinfo {volume}
  {97}},\ \bibinfo {pages} {045504} (\bibinfo {year} {2006})}\BibitemShut
  {NoStop}%
\bibitem [{\citenamefont {Wang}\ \emph {et~al.}(2010)\citenamefont {Wang},
  \citenamefont {Lv}, \citenamefont {Zhu},\ and\ \citenamefont
  {Ma}}]{wang2010crystal}%
  \BibitemOpen
  \bibfield  {author} {\bibinfo {author} {\bibfnamefont {Y.}~\bibnamefont
  {Wang}}, \bibinfo {author} {\bibfnamefont {J.}~\bibnamefont {Lv}}, \bibinfo
  {author} {\bibfnamefont {L.}~\bibnamefont {Zhu}}, \ and\ \bibinfo {author}
  {\bibfnamefont {Y.}~\bibnamefont {Ma}},\ }\href {\doibase
  10.1103/PhysRevB.82.094116} {\bibfield  {journal} {\bibinfo  {journal}
  {Phys.Rev.B}\ }\textbf {\bibinfo {volume} {82}},\ \bibinfo {pages} {094116}
  (\bibinfo {year} {2010})}\BibitemShut {NoStop}%
\bibitem [{\citenamefont {Pickard}\ \emph {et~al.}(2012)\citenamefont
  {Pickard}, \citenamefont {Martinez-Canales},\ and\ \citenamefont
  {Needs}}]{pickard2012density}%
  \BibitemOpen
  \bibfield  {author} {\bibinfo {author} {\bibfnamefont {C.~J.}\ \bibnamefont
  {Pickard}}, \bibinfo {author} {\bibfnamefont {M.}~\bibnamefont
  {Martinez-Canales}}, \ and\ \bibinfo {author} {\bibfnamefont {R.~J.}\
  \bibnamefont {Needs}},\ }\href {\doibase 10.1103/PhysRevB.85.214114}
  {\bibfield  {journal} {\bibinfo  {journal} {Phys.Rev.B}\ }\textbf {\bibinfo
  {volume} {85}},\ \bibinfo {pages} {214114} (\bibinfo {year}
  {2012})}\BibitemShut {NoStop}%
\bibitem [{\citenamefont {Liu}\ \emph {et~al.}(2012)\citenamefont {Liu},
  \citenamefont {Wang},\ and\ \citenamefont {Ma}}]{liu2012quasi}%
  \BibitemOpen
  \bibfield  {author} {\bibinfo {author} {\bibfnamefont {H.}~\bibnamefont
  {Liu}}, \bibinfo {author} {\bibfnamefont {H.}~\bibnamefont {Wang}}, \ and\
  \bibinfo {author} {\bibfnamefont {Y.}~\bibnamefont {Ma}},\ }\href {\doibase
  10.1021/jp301596} {\bibfield  {journal} {\bibinfo  {journal} {J.Phys.Chem.C}\
  }\textbf {\bibinfo {volume} {116}},\ \bibinfo {pages} {9221} (\bibinfo {year}
  {2012})}\BibitemShut {NoStop}%
\bibitem [{\citenamefont {Eremets}\ and\ \citenamefont
  {Drozdov}(2017)}]{eremets2017comments}%
  \BibitemOpen
  \bibfield  {author} {\bibinfo {author} {\bibfnamefont {M.~I.}\ \bibnamefont
  {Eremets}}\ and\ \bibinfo {author} {\bibfnamefont {A.~P.}\ \bibnamefont
  {Drozdov}},\ }\href@noop {} {\bibfield  {journal} {\bibinfo  {journal}
  {arXiv:1702.05125}\ } (\bibinfo {year} {2017})}\BibitemShut {NoStop}%
\bibitem [{\citenamefont {Loubeyre}\ \emph {et~al.}()\citenamefont {Loubeyre},
  \citenamefont {Occelli},\ and\ \citenamefont {Dumas}}]{loubeyre1702comment}%
  \BibitemOpen
  \bibfield  {author} {\bibinfo {author} {\bibfnamefont {P.}~\bibnamefont
  {Loubeyre}}, \bibinfo {author} {\bibfnamefont {F.}~\bibnamefont {Occelli}}, \
  and\ \bibinfo {author} {\bibfnamefont {P.}~\bibnamefont {Dumas}},\
  }\href@noop {} {\bibinfo  {journal} {arXiv:1702.07192}\ }\BibitemShut
  {NoStop}%
\bibitem [{\citenamefont {Silvera}\ and\ \citenamefont
  {Dias}(2017{\natexlab{a}})}]{silvera2017response}%
  \BibitemOpen
\bibfield  {journal} {  }\bibfield  {author} {\bibinfo {author} {\bibfnamefont
  {I.}~\bibnamefont {Silvera}}\ and\ \bibinfo {author} {\bibfnamefont
  {R.}~\bibnamefont {Dias}},\ }\href@noop {} {\bibfield  {journal} {\bibinfo
  {journal} {arXiv:1703.03064}\ } (\bibinfo {year}
  {2017}{\natexlab{a}})}\BibitemShut {NoStop}%
\bibitem [{\citenamefont {Goncharov}\ and\ \citenamefont
  {Struzhkin}(2017)}]{goncharov2017comment}%
  \BibitemOpen
  \bibfield  {author} {\bibinfo {author} {\bibfnamefont {A.~F.}\ \bibnamefont
  {Goncharov}}\ and\ \bibinfo {author} {\bibfnamefont {V.~V.}\ \bibnamefont
  {Struzhkin}},\ }\href {\doibase 10.1126/science.aam9736} {\bibfield
  {journal} {\bibinfo  {journal} {Science}\ }\textbf {\bibinfo {volume}
  {357}},\ \bibinfo {pages} {eaam9736} (\bibinfo {year} {2017})}\BibitemShut
  {NoStop}%
\bibitem [{\citenamefont {Liu}\ \emph {et~al.}(2017)\citenamefont {Liu},
  \citenamefont {Dalladay-Simpson}, \citenamefont {Howie}, \citenamefont {Li},\
  and\ \citenamefont {Gregoryanz}}]{liu2017comment}%
  \BibitemOpen
  \bibfield  {author} {\bibinfo {author} {\bibfnamefont {X.-D.}\ \bibnamefont
  {Liu}}, \bibinfo {author} {\bibfnamefont {P.}~\bibnamefont
  {Dalladay-Simpson}}, \bibinfo {author} {\bibfnamefont {R.~T.}\ \bibnamefont
  {Howie}}, \bibinfo {author} {\bibfnamefont {B.}~\bibnamefont {Li}}, \ and\
  \bibinfo {author} {\bibfnamefont {E.}~\bibnamefont {Gregoryanz}},\ }\href
  {\doibase 10.1126/science.aan2286} {\bibfield  {journal} {\bibinfo  {journal}
  {Science}\ }\textbf {\bibinfo {volume} {357}},\ \bibinfo {pages} {eaan2286}
  (\bibinfo {year} {2017})}\BibitemShut {NoStop}%
\bibitem [{\citenamefont {Silvera}\ and\ \citenamefont
  {Dias}(2017{\natexlab{b}})}]{silvera2017science}%
  \BibitemOpen
  \bibfield  {author} {\bibinfo {author} {\bibfnamefont {I.~F.}\ \bibnamefont
  {Silvera}}\ and\ \bibinfo {author} {\bibfnamefont {R.}~\bibnamefont {Dias}},\
  }\href {\doibase 10.1126/science.aan2671} {\bibfield  {journal} {\bibinfo
  {journal} {Science}\ }\textbf {\bibinfo {volume} {357}},\ \bibinfo {pages}
  {eaan2671} (\bibinfo {year} {2017}{\natexlab{b}})}\BibitemShut {NoStop}%
\bibitem [{\citenamefont {Geng}(2017)}]{geng2017public}%
  \BibitemOpen
  \bibfield  {author} {\bibinfo {author} {\bibfnamefont {H.~Y.}\ \bibnamefont
  {Geng}},\ }\href {\doibase 10.1016/j.mre.2017.10.001} {\bibfield  {journal}
  {\bibinfo  {journal} {Matter Radiat.Extremes}\ }\textbf {\bibinfo {volume}
  {2}},\ \bibinfo {pages} {275} (\bibinfo {year} {2017})}\BibitemShut {NoStop}%
\bibitem [{\citenamefont {Borinaga}\ \emph {et~al.}(2018)\citenamefont
  {Borinaga}, \citenamefont {Iba{\~n}ez-Azpiroz}, \citenamefont {Bergara},\
  and\ \citenamefont {Errea}}]{borinaga2018strong}%
  \BibitemOpen
  \bibfield  {author} {\bibinfo {author} {\bibfnamefont {M.}~\bibnamefont
  {Borinaga}}, \bibinfo {author} {\bibfnamefont {J.}~\bibnamefont
  {Iba{\~n}ez-Azpiroz}}, \bibinfo {author} {\bibfnamefont {A.}~\bibnamefont
  {Bergara}}, \ and\ \bibinfo {author} {\bibfnamefont {I.}~\bibnamefont
  {Errea}},\ }\href {\doibase 10.1103/PhysRevLett.120.057402} {\bibfield
  {journal} {\bibinfo  {journal} {Phys.Rev.Lett.}\ }\textbf {\bibinfo {volume}
  {120}},\ \bibinfo {pages} {057402} (\bibinfo {year} {2018})}\BibitemShut
  {NoStop}%
\bibitem [{\citenamefont {Ambrosch-Draxl}\ and\ \citenamefont
  {Sofo}(2006)}]{ambrosch2006linear}%
  \BibitemOpen
  \bibfield  {author} {\bibinfo {author} {\bibfnamefont {C.}~\bibnamefont
  {Ambrosch-Draxl}}\ and\ \bibinfo {author} {\bibfnamefont {J.~O.}\
  \bibnamefont {Sofo}},\ }\href@noop {} {\bibfield  {journal} {\bibinfo
  {journal} {‎Comput. Phys. Commun.}\ }\textbf {\bibinfo {volume} {175}},\
  \bibinfo {pages} {1} (\bibinfo {year} {2006})}\BibitemShut {NoStop}%
\bibitem [{\citenamefont {Gajdo{\v{s}}}\ \emph {et~al.}(2006)\citenamefont
  {Gajdo{\v{s}}}, \citenamefont {Hummer}, \citenamefont {Kresse}, \citenamefont
  {Furthm{\"u}ller},\ and\ \citenamefont {Bechstedt}}]{gajdovs2006linear}%
  \BibitemOpen
  \bibfield  {author} {\bibinfo {author} {\bibfnamefont {M.}~\bibnamefont
  {Gajdo{\v{s}}}}, \bibinfo {author} {\bibfnamefont {K.}~\bibnamefont
  {Hummer}}, \bibinfo {author} {\bibfnamefont {G.}~\bibnamefont {Kresse}},
  \bibinfo {author} {\bibfnamefont {J.}~\bibnamefont {Furthm{\"u}ller}}, \ and\
  \bibinfo {author} {\bibfnamefont {F.}~\bibnamefont {Bechstedt}},\ }\href
  {\doibase 10.1103/PhysRevB.73.045112} {\bibfield  {journal} {\bibinfo
  {journal} {Phys.Rev.B}\ }\textbf {\bibinfo {volume} {73}},\ \bibinfo {pages}
  {045112} (\bibinfo {year} {2006})}\BibitemShut {NoStop}%
\bibitem [{\citenamefont {Aryasetiawan}\ and\ \citenamefont
  {Gunnarsson}(1998)}]{aryasetiawan1998gw}%
  \BibitemOpen
  \bibfield  {author} {\bibinfo {author} {\bibfnamefont {F.}~\bibnamefont
  {Aryasetiawan}}\ and\ \bibinfo {author} {\bibfnamefont {O.}~\bibnamefont
  {Gunnarsson}},\ }\href {\doibase 10.1088/0034-4885/61/3/002} {\bibfield
  {journal} {\bibinfo  {journal} {‎Rep. Prog. Phys.}\ }\textbf {\bibinfo
  {volume} {61}},\ \bibinfo {pages} {237} (\bibinfo {year} {1998})}\BibitemShut
  {NoStop}%
\bibitem [{\citenamefont {Louie}\ \emph {et~al.}(1975)\citenamefont {Louie},
  \citenamefont {Chelikowsky},\ and\ \citenamefont {Cohen}}]{louie1975}%
  \BibitemOpen
  \bibfield  {author} {\bibinfo {author} {\bibfnamefont {S.~G.}\ \bibnamefont
  {Louie}}, \bibinfo {author} {\bibfnamefont {J.~R.}\ \bibnamefont
  {Chelikowsky}}, \ and\ \bibinfo {author} {\bibfnamefont {M.~L.}\ \bibnamefont
  {Cohen}},\ }\href@noop {} {\bibfield  {journal} {\bibinfo  {journal} {Phys.
  Rev. Lett.}\ }\textbf {\bibinfo {volume} {34}},\ \bibinfo {pages} {155}
  (\bibinfo {year} {1975})}\BibitemShut {NoStop}%
\bibitem [{\citenamefont {Harl}(2008)}]{harl2008linear}%
  \BibitemOpen
  \bibfield  {author} {\bibinfo {author} {\bibfnamefont {J.}~\bibnamefont
  {Harl}},\ }\emph {\bibinfo {title} {The linear response function in density
  functional theory}},\ \href {http://othes.univie.ac.at/2622/} {Ph.D.
  thesis},\ \bibinfo  {school} {University of Vienna} (\bibinfo {year}
  {2008})\BibitemShut {NoStop}%
\bibitem [{\citenamefont {Lax}(1952)}]{lax1952franck}%
  \BibitemOpen
  \bibfield  {author} {\bibinfo {author} {\bibfnamefont {M.}~\bibnamefont
  {Lax}},\ }\href {\doibase 10.1063/1.1700283} {\bibfield  {journal} {\bibinfo
  {journal} {J.Chem.Phys.}\ }\textbf {\bibinfo {volume} {20}},\ \bibinfo
  {pages} {1752} (\bibinfo {year} {1952})}\BibitemShut {NoStop}%
\bibitem [{\citenamefont {Patrick}\ and\ \citenamefont
  {Giustino}(2014)}]{patrick2014unified}%
  \BibitemOpen
  \bibfield  {author} {\bibinfo {author} {\bibfnamefont {C.~E.}\ \bibnamefont
  {Patrick}}\ and\ \bibinfo {author} {\bibfnamefont {F.}~\bibnamefont
  {Giustino}},\ }\href {\doibase 10.1088/0953-8984/26/36/365503} {\bibfield
  {journal} {\bibinfo  {journal} {‎J. Phys. Condens. Matter}\ }\textbf
  {\bibinfo {volume} {26}},\ \bibinfo {pages} {365503} (\bibinfo {year}
  {2014})}\BibitemShut {NoStop}%
\bibitem [{\citenamefont {Della~Sala}\ \emph {et~al.}(2004)\citenamefont
  {Della~Sala}, \citenamefont {Rousseau}, \citenamefont {G{\"o}rling},\ and\
  \citenamefont {Marx}}]{della2004quantum}%
  \BibitemOpen
  \bibfield  {author} {\bibinfo {author} {\bibfnamefont {F.}~\bibnamefont
  {Della~Sala}}, \bibinfo {author} {\bibfnamefont {R.}~\bibnamefont
  {Rousseau}}, \bibinfo {author} {\bibfnamefont {A.}~\bibnamefont
  {G{\"o}rling}}, \ and\ \bibinfo {author} {\bibfnamefont {D.}~\bibnamefont
  {Marx}},\ }\href {\doibase 10.1103/PhysRevLett.92.183401} {\bibfield
  {journal} {\bibinfo  {journal} {Phy.Rev.Lett.}\ }\textbf {\bibinfo {volume}
  {92}},\ \bibinfo {pages} {183401} (\bibinfo {year} {2004})}\BibitemShut
  {NoStop}%
\bibitem [{\citenamefont {Tuckerman}(2010)}]{tuckerman2010statistical}%
  \BibitemOpen
  \bibfield  {author} {\bibinfo {author} {\bibfnamefont {M.}~\bibnamefont
  {Tuckerman}},\ }\href@noop {} {\emph {\bibinfo {title} {Statistical
  mechanics: theory and molecular simulation}}}\ (\bibinfo  {publisher} {Oxford
  University Press},\ \bibinfo {year} {2010})\BibitemShut {NoStop}%
\bibitem [{\citenamefont {Li}\ and\ \citenamefont
  {Wang}(2018)}]{enge2018computer}%
  \BibitemOpen
  \bibfield  {author} {\bibinfo {author} {\bibfnamefont {X.~Z.}\ \bibnamefont
  {Li}}\ and\ \bibinfo {author} {\bibfnamefont {E.~G.}\ \bibnamefont {Wang}},\
  }\href@noop {} {\emph {\bibinfo {title} {Computer Simulations Of Molecules
  And Condensed Matter: From Electronic Structures To Molecular Dynamics}}}\
  (\bibinfo  {publisher} {World Scientific},\ \bibinfo {year}
  {2018})\BibitemShut {NoStop}%
\bibitem [{\citenamefont {Zacharias}\ and\ \citenamefont
  {Giustino}(2016)}]{zacharias2016one}%
  \BibitemOpen
  \bibfield  {author} {\bibinfo {author} {\bibfnamefont {M.}~\bibnamefont
  {Zacharias}}\ and\ \bibinfo {author} {\bibfnamefont {F.}~\bibnamefont
  {Giustino}},\ }\href {\doibase 10.1103/PhysRevB.94.075125} {\bibfield
  {journal} {\bibinfo  {journal} {Phys.Rev.B}\ }\textbf {\bibinfo {volume}
  {94}},\ \bibinfo {pages} {075125} (\bibinfo {year} {2016})}\BibitemShut
  {NoStop}%
\bibitem [{\citenamefont {Zacharias}\ \emph {et~al.}(2015)\citenamefont
  {Zacharias}, \citenamefont {Patrick},\ and\ \citenamefont
  {Giustino}}]{zacharias2015stochastic}%
  \BibitemOpen
  \bibfield  {author} {\bibinfo {author} {\bibfnamefont {M.}~\bibnamefont
  {Zacharias}}, \bibinfo {author} {\bibfnamefont {C.~E.}\ \bibnamefont
  {Patrick}}, \ and\ \bibinfo {author} {\bibfnamefont {F.}~\bibnamefont
  {Giustino}},\ }\href {\doibase 10.1103/PhysRevLett.115.177401} {\bibfield
  {journal} {\bibinfo  {journal} {Phys.Rev.Lett.}\ }\textbf {\bibinfo {volume}
  {115}},\ \bibinfo {pages} {177401} (\bibinfo {year} {2015})}\BibitemShut
  {NoStop}%
\bibitem [{\citenamefont {Marx}\ and\ \citenamefont
  {Parrinello}(1994)}]{marx1994ab}%
  \BibitemOpen
  \bibfield  {author} {\bibinfo {author} {\bibfnamefont {D.}~\bibnamefont
  {Marx}}\ and\ \bibinfo {author} {\bibfnamefont {M.}~\bibnamefont
  {Parrinello}},\ }\href {\doibase 10.1007/BF01312185} {\bibfield  {journal}
  {\bibinfo  {journal} {Z.Phys.B}\ }\textbf {\bibinfo {volume} {95}},\ \bibinfo
  {pages} {143} (\bibinfo {year} {1994})}\BibitemShut {NoStop}%
\bibitem [{\citenamefont {Marx}\ and\ \citenamefont
  {Parrinello}(1996)}]{marx1996ab}%
  \BibitemOpen
  \bibfield  {author} {\bibinfo {author} {\bibfnamefont {D.}~\bibnamefont
  {Marx}}\ and\ \bibinfo {author} {\bibfnamefont {M.}~\bibnamefont
  {Parrinello}},\ }\href {\doibase 10.1063/1.471221} {\bibfield  {journal}
  {\bibinfo  {journal} {J.Chem.Phys.}\ }\textbf {\bibinfo {volume} {104}},\
  \bibinfo {pages} {4077} (\bibinfo {year} {1996})}\BibitemShut {NoStop}%
\bibitem [{\citenamefont {Tuckerman}\ \emph {et~al.}(1996)\citenamefont
  {Tuckerman}, \citenamefont {Marx}, \citenamefont {Klein},\ and\ \citenamefont
  {Parrinello}}]{tuckerman1996efficient}%
  \BibitemOpen
  \bibfield  {author} {\bibinfo {author} {\bibfnamefont {M.~E.}\ \bibnamefont
  {Tuckerman}}, \bibinfo {author} {\bibfnamefont {D.}~\bibnamefont {Marx}},
  \bibinfo {author} {\bibfnamefont {M.~L.}\ \bibnamefont {Klein}}, \ and\
  \bibinfo {author} {\bibfnamefont {M.}~\bibnamefont {Parrinello}},\ }\href
  {\doibase 10.1063/1.471771} {\bibfield  {journal} {\bibinfo  {journal}
  {J.Chem.Phys.}\ }\textbf {\bibinfo {volume} {104}},\ \bibinfo {pages} {5579}
  (\bibinfo {year} {1996})}\BibitemShut {NoStop}%
\bibitem [{\citenamefont {Kresse}\ and\ \citenamefont
  {Furthm{\"u}ller}(1996{\natexlab{a}})}]{kresse1996efficiency}%
  \BibitemOpen
  \bibfield  {author} {\bibinfo {author} {\bibfnamefont {G.}~\bibnamefont
  {Kresse}}\ and\ \bibinfo {author} {\bibfnamefont {J.}~\bibnamefont
  {Furthm{\"u}ller}},\ }\href {\doibase 10.1016/0927-0256(96)00008-0}
  {\bibfield  {journal} {\bibinfo  {journal} {Comput. Mater. Sci.}\ }\textbf
  {\bibinfo {volume} {6}},\ \bibinfo {pages} {15} (\bibinfo {year}
  {1996}{\natexlab{a}})}\BibitemShut {NoStop}%
\bibitem [{\citenamefont {Kresse}\ and\ \citenamefont
  {Furthm{\"u}ller}(1996{\natexlab{b}})}]{kresse1996efficient}%
  \BibitemOpen
  \bibfield  {author} {\bibinfo {author} {\bibfnamefont {G.}~\bibnamefont
  {Kresse}}\ and\ \bibinfo {author} {\bibfnamefont {J.}~\bibnamefont
  {Furthm{\"u}ller}},\ }\href {\doibase 10.1103/PhysRevB.54.11169} {\bibfield
  {journal} {\bibinfo  {journal} {Phys.Rev.B}\ }\textbf {\bibinfo {volume}
  {54}},\ \bibinfo {pages} {11169} (\bibinfo {year}
  {1996}{\natexlab{b}})}\BibitemShut {NoStop}%
\bibitem [{\citenamefont {Bl{\"o}chl}(1994)}]{blochl1994projector}%
  \BibitemOpen
  \bibfield  {author} {\bibinfo {author} {\bibfnamefont {P.~E.}\ \bibnamefont
  {Bl{\"o}chl}},\ }\href {\doibase 10.1103/PhysRevB.50.17953} {\bibfield
  {journal} {\bibinfo  {journal} {Phys.Rev.B}\ }\textbf {\bibinfo {volume}
  {50}},\ \bibinfo {pages} {17953} (\bibinfo {year} {1994})}\BibitemShut
  {NoStop}%
\bibitem [{\citenamefont {Kresse}\ and\ \citenamefont
  {Joubert}(1999)}]{kresse1999ultrasoft}%
  \BibitemOpen
  \bibfield  {author} {\bibinfo {author} {\bibfnamefont {G.}~\bibnamefont
  {Kresse}}\ and\ \bibinfo {author} {\bibfnamefont {D.}~\bibnamefont
  {Joubert}},\ }\href {\doibase 10.1103/PhysRevB.59.1758} {\bibfield  {journal}
  {\bibinfo  {journal} {Phys.Rev.B}\ }\textbf {\bibinfo {volume} {59}},\
  \bibinfo {pages} {1758} (\bibinfo {year} {1999})}\BibitemShut {NoStop}%
\bibitem [{\citenamefont {Perdew}\ \emph {et~al.}(1996)\citenamefont {Perdew},
  \citenamefont {Burke},\ and\ \citenamefont
  {Ernzerhof}}]{perdew1996generalized}%
  \BibitemOpen
  \bibfield  {author} {\bibinfo {author} {\bibfnamefont {J.~P.}\ \bibnamefont
  {Perdew}}, \bibinfo {author} {\bibfnamefont {K.}~\bibnamefont {Burke}}, \
  and\ \bibinfo {author} {\bibfnamefont {M.}~\bibnamefont {Ernzerhof}},\ }\href
  {\doibase 10.1103/PhysRevLett.77.3865} {\bibfield  {journal} {\bibinfo
  {journal} {Phys.Rev.Lett.}\ }\textbf {\bibinfo {volume} {77}},\ \bibinfo
  {pages} {3865} (\bibinfo {year} {1996})}\BibitemShut {NoStop}%
\bibitem [{\citenamefont {Togo}\ and\ \citenamefont
  {Tanaka}(2015)}]{togo2015first}%
  \BibitemOpen
  \bibfield  {author} {\bibinfo {author} {\bibfnamefont {A.}~\bibnamefont
  {Togo}}\ and\ \bibinfo {author} {\bibfnamefont {I.}~\bibnamefont {Tanaka}},\
  }\href {\doibase 10.1016/j.scriptamat.2015.07.021} {\bibfield  {journal}
  {\bibinfo  {journal} {Scr. Mater.}\ }\textbf {\bibinfo {volume} {108}},\
  \bibinfo {pages} {1} (\bibinfo {year} {2015})}\BibitemShut {NoStop}%
\bibitem [{\citenamefont {Baroni}\ \emph {et~al.}(2001)\citenamefont {Baroni},
  \citenamefont {De~Gironcoli}, \citenamefont {Dal~Corso},\ and\ \citenamefont
  {Giannozzi}}]{baroni2001phonons}%
  \BibitemOpen
  \bibfield  {author} {\bibinfo {author} {\bibfnamefont {S.}~\bibnamefont
  {Baroni}}, \bibinfo {author} {\bibfnamefont {S.}~\bibnamefont
  {De~Gironcoli}}, \bibinfo {author} {\bibfnamefont {A.}~\bibnamefont
  {Dal~Corso}}, \ and\ \bibinfo {author} {\bibfnamefont {P.}~\bibnamefont
  {Giannozzi}},\ }\href {\doibase 10.1103/RevModPhys.73.515} {\bibfield
  {journal} {\bibinfo  {journal} {Rev.Mod.Phys.}\ }\textbf {\bibinfo {volume}
  {73}},\ \bibinfo {pages} {515} (\bibinfo {year} {2001})}\BibitemShut
  {NoStop}%
\bibitem [{\citenamefont {Giannozzi}\ \emph {et~al.}(2009)\citenamefont
  {Giannozzi}, \citenamefont {Baroni}, \citenamefont {Bonini}, \citenamefont
  {Calandra}, \citenamefont {Car}, \citenamefont {Cavazzoni}, \citenamefont
  {Ceresoli}, \citenamefont {Chiarotti}, \citenamefont {Cococcioni},
  \citenamefont {Dabo} \emph {et~al.}}]{giannozzi2009quantum}%
  \BibitemOpen
  \bibfield  {author} {\bibinfo {author} {\bibfnamefont {P.}~\bibnamefont
  {Giannozzi}}, \bibinfo {author} {\bibfnamefont {S.}~\bibnamefont {Baroni}},
  \bibinfo {author} {\bibfnamefont {N.}~\bibnamefont {Bonini}}, \bibinfo
  {author} {\bibfnamefont {M.}~\bibnamefont {Calandra}}, \bibinfo {author}
  {\bibfnamefont {R.}~\bibnamefont {Car}}, \bibinfo {author} {\bibfnamefont
  {C.}~\bibnamefont {Cavazzoni}}, \bibinfo {author} {\bibfnamefont
  {D.}~\bibnamefont {Ceresoli}}, \bibinfo {author} {\bibfnamefont {G.~L.}\
  \bibnamefont {Chiarotti}}, \bibinfo {author} {\bibfnamefont {M.}~\bibnamefont
  {Cococcioni}}, \bibinfo {author} {\bibfnamefont {I.}~\bibnamefont {Dabo}},
  \emph {et~al.},\ }\href {\doibase 10.1088/0953-8984/21/39/395502} {\bibfield
  {journal} {\bibinfo  {journal} {J. Phys.: Condens. Matter}\ }\textbf
  {\bibinfo {volume} {21}},\ \bibinfo {pages} {395502} (\bibinfo {year}
  {2009})}\BibitemShut {NoStop}%
\bibitem [{\citenamefont {Hedin}(1965)}]{hedin1965}%
  \BibitemOpen
  \bibfield  {author} {\bibinfo {author} {\bibfnamefont {L.}~\bibnamefont
  {Hedin}},\ }\href@noop {} {\bibfield  {journal} {\bibinfo  {journal} {Phys.
  Rev.}\ }\textbf {\bibinfo {volume} {139}},\ \bibinfo {pages} {A796} (\bibinfo
  {year} {1965})}\BibitemShut {NoStop}%
\bibitem [{\citenamefont {Hybertsen}\ and\ \citenamefont
  {Louie}(1986)}]{hybertsen1986}%
  \BibitemOpen
  \bibfield  {author} {\bibinfo {author} {\bibfnamefont {M.~S.}\ \bibnamefont
  {Hybertsen}}\ and\ \bibinfo {author} {\bibfnamefont {S.~G.}\ \bibnamefont
  {Louie}},\ }\href@noop {} {\bibfield  {journal} {\bibinfo  {journal} {Phys.
  Rev. B}\ }\textbf {\bibinfo {volume} {34}},\ \bibinfo {pages} {5390}
  (\bibinfo {year} {1986})}\BibitemShut {NoStop}%
\bibitem [{\citenamefont {Shishkin}\ and\ \citenamefont
  {Kresse}(2006)}]{shishkin2006}%
  \BibitemOpen
  \bibfield  {author} {\bibinfo {author} {\bibfnamefont {M.}~\bibnamefont
  {Shishkin}}\ and\ \bibinfo {author} {\bibfnamefont {G.}~\bibnamefont
  {Kresse}},\ }\href@noop {} {\bibfield  {journal} {\bibinfo  {journal} {Phys.
  Rev. B}\ }\textbf {\bibinfo {volume} {74}},\ \bibinfo {pages} {035101}
  (\bibinfo {year} {2006})}\BibitemShut {NoStop}%
\bibitem [{\citenamefont {Shishkin}\ and\ \citenamefont
  {Kresse}(2007)}]{shishkin2007}%
  \BibitemOpen
  \bibfield  {author} {\bibinfo {author} {\bibfnamefont {M.}~\bibnamefont
  {Shishkin}}\ and\ \bibinfo {author} {\bibfnamefont {G.}~\bibnamefont
  {Kresse}},\ }\href@noop {} {\bibfield  {journal} {\bibinfo  {journal} {Phys.
  Rev. B}\ }\textbf {\bibinfo {volume} {75}},\ \bibinfo {pages} {235102}
  (\bibinfo {year} {2007})}\BibitemShut {NoStop}%
\bibitem [{\citenamefont {Li}\ \emph {et~al.}(2013)\citenamefont {Li},
  \citenamefont {Probert}, \citenamefont {Pickard}, \citenamefont {Needs},\
  and\ \citenamefont {Michaelides}}]{Li2013}%
  \BibitemOpen
  \bibfield  {author} {\bibinfo {author} {\bibfnamefont {X.~Z.}\ \bibnamefont
  {Li}}, \bibinfo {author} {\bibfnamefont {M.~I.~J.}\ \bibnamefont {Probert}},
  \bibinfo {author} {\bibfnamefont {C.~J.}\ \bibnamefont {Pickard}}, \bibinfo
  {author} {\bibfnamefont {R.~J.}\ \bibnamefont {Needs}}, \ and\ \bibinfo
  {author} {\bibfnamefont {A.}~\bibnamefont {Michaelides}},\ }\href {\doibase
  10.1088/0953-8984/25/8/085402} {\bibfield  {journal} {\bibinfo  {journal} {J.
  Phys.: Condens. Matter}\ }\textbf {\bibinfo {volume} {25}},\ \bibinfo {pages}
  {085402} (\bibinfo {year} {2013})}\BibitemShut {NoStop}%
\bibitem [{\citenamefont {Drummond}\ \emph {et~al.}(2015)\citenamefont
  {Drummond}, \citenamefont {Monserrat}, \citenamefont {Lloyd-Williams},
  \citenamefont {R{\'\i}os}, \citenamefont {Pickard},\ and\ \citenamefont
  {Needs}}]{drummond2015quantum}%
  \BibitemOpen
  \bibfield  {author} {\bibinfo {author} {\bibfnamefont {N.~D.}\ \bibnamefont
  {Drummond}}, \bibinfo {author} {\bibfnamefont {B.}~\bibnamefont {Monserrat}},
  \bibinfo {author} {\bibfnamefont {J.~H.}\ \bibnamefont {Lloyd-Williams}},
  \bibinfo {author} {\bibfnamefont {P.~L.}\ \bibnamefont {R{\'\i}os}}, \bibinfo
  {author} {\bibfnamefont {C.~J.}\ \bibnamefont {Pickard}}, \ and\ \bibinfo
  {author} {\bibfnamefont {R.~J.}\ \bibnamefont {Needs}},\ }\href {\doibase
  10.1038/ncomms8794} {\bibfield  {journal} {\bibinfo  {journal} {Nat.Commun.}\
  }\textbf {\bibinfo {volume} {6}},\ \bibinfo {pages} {7794} (\bibinfo {year}
  {2015})}\BibitemShut {NoStop}%
\bibitem [{\citenamefont {Azadi}\ and\ \citenamefont
  {Foulkes}(2013)}]{azadi2013fate}%
  \BibitemOpen
  \bibfield  {author} {\bibinfo {author} {\bibfnamefont {S.}~\bibnamefont
  {Azadi}}\ and\ \bibinfo {author} {\bibfnamefont {W.~M.~C.}\ \bibnamefont
  {Foulkes}},\ }\href {\doibase 10.1103/PhysRevB.88.014115} {\bibfield
  {journal} {\bibinfo  {journal} {Phys.Rev.B}\ }\textbf {\bibinfo {volume}
  {88}},\ \bibinfo {pages} {014115} (\bibinfo {year} {2013})}\BibitemShut
  {NoStop}%
\bibitem [{\citenamefont {Marini}(2001)}]{marini2001optical}%
  \BibitemOpen
  \bibfield  {author} {\bibinfo {author} {\bibfnamefont {A.}~\bibnamefont
  {Marini}},\ }\emph {\bibinfo {title} {Optical and electronic properties of
  copper and silver: From density functional theory to many body effects}},\
  \href
  {http://www.yambo-code.org/people/andrea/wp-content/uploads/2015/06/thesis.pdf}
  {Ph.D. thesis},\ \bibinfo  {school} {Ph. D. dissertation (University of Rome
  Tor Vergata)} (\bibinfo {year} {2001})\BibitemShut {NoStop}%
\bibitem [{\citenamefont {Medeiros}\ \emph {et~al.}(2014)\citenamefont
  {Medeiros}, \citenamefont {Stafstr{\"o}m},\ and\ \citenamefont
  {Bj{\"o}rk}}]{medeiros2014effects}%
  \BibitemOpen
  \bibfield  {author} {\bibinfo {author} {\bibfnamefont {P.~V.~C.}\
  \bibnamefont {Medeiros}}, \bibinfo {author} {\bibfnamefont {S.}~\bibnamefont
  {Stafstr{\"o}m}}, \ and\ \bibinfo {author} {\bibfnamefont {J.}~\bibnamefont
  {Bj{\"o}rk}},\ }\href {\doibase 10.1103/PhysRevB.89.041407} {\bibfield
  {journal} {\bibinfo  {journal} {Phys.Rev.B}\ }\textbf {\bibinfo {volume}
  {89}},\ \bibinfo {pages} {041407} (\bibinfo {year} {2014})}\BibitemShut
  {NoStop}%
\bibitem [{\citenamefont {Medeiros}\ \emph {et~al.}(2015)\citenamefont
  {Medeiros}, \citenamefont {Tsirkin}, \citenamefont {Stafstr{\"o}m},\ and\
  \citenamefont {Bj{\"o}rk}}]{medeiros2015unfolding}%
  \BibitemOpen
  \bibfield  {author} {\bibinfo {author} {\bibfnamefont {P.~V.~C.}\
  \bibnamefont {Medeiros}}, \bibinfo {author} {\bibfnamefont {S.~S.}\
  \bibnamefont {Tsirkin}}, \bibinfo {author} {\bibfnamefont {S.}~\bibnamefont
  {Stafstr{\"o}m}}, \ and\ \bibinfo {author} {\bibfnamefont {J.}~\bibnamefont
  {Bj{\"o}rk}},\ }\href {\doibase 10.1103/PhysRevB.91.041116} {\bibfield
  {journal} {\bibinfo  {journal} {Phys.Rev.B}\ }\textbf {\bibinfo {volume}
  {91}},\ \bibinfo {pages} {041116} (\bibinfo {year} {2015})}\BibitemShut
  {NoStop}%
\bibitem [{\citenamefont {Surh}\ \emph {et~al.}(1992)\citenamefont {Surh},
  \citenamefont {Louie},\ and\ \citenamefont {Cohen}}]{surh1992band}%
  \BibitemOpen
  \bibfield  {author} {\bibinfo {author} {\bibfnamefont {M.~P.}\ \bibnamefont
  {Surh}}, \bibinfo {author} {\bibfnamefont {S.~G.}\ \bibnamefont {Louie}}, \
  and\ \bibinfo {author} {\bibfnamefont {M.~L.}\ \bibnamefont {Cohen}},\ }\href
  {\doibase 10.1103/PhysRevB.45.8239} {\bibfield  {journal} {\bibinfo
  {journal} {Phys.Rev.B}\ }\textbf {\bibinfo {volume} {45}},\ \bibinfo {pages}
  {8239} (\bibinfo {year} {1992})}\BibitemShut {NoStop}%
\bibitem [{\citenamefont {Gomez-Abal}\ \emph {et~al.}(2008)\citenamefont
  {Gomez-Abal}, \citenamefont {Li}, \citenamefont {Ambrosh-Draxl},\ and\
  \citenamefont {Scheffler}}]{ricardo2008}%
  \BibitemOpen
  \bibfield  {author} {\bibinfo {author} {\bibfnamefont {R.~I.}\ \bibnamefont
  {Gomez-Abal}}, \bibinfo {author} {\bibfnamefont {X.~Z.}\ \bibnamefont {Li}},
  \bibinfo {author} {\bibfnamefont {C.}~\bibnamefont {Ambrosh-Draxl}}, \ and\
  \bibinfo {author} {\bibfnamefont {M.}~\bibnamefont {Scheffler}},\ }\href@noop
  {} {\bibfield  {journal} {\bibinfo  {journal} {Phys. Rev. Lett.}\ }\textbf
  {\bibinfo {volume} {101}},\ \bibinfo {pages} {106404} (\bibinfo {year}
  {2008})}\BibitemShut {NoStop}%
\bibitem [{\citenamefont {Li}\ \emph {et~al.}(2012)\citenamefont {Li},
  \citenamefont {Gomez-Abal}, \citenamefont {Jiang}, \citenamefont
  {Ambrosh-Draxl},\ and\ \citenamefont {Scheffler}}]{li2012}%
  \BibitemOpen
  \bibfield  {author} {\bibinfo {author} {\bibfnamefont {X.~Z.}\ \bibnamefont
  {Li}}, \bibinfo {author} {\bibfnamefont {R.~I.}\ \bibnamefont {Gomez-Abal}},
  \bibinfo {author} {\bibfnamefont {H.}~\bibnamefont {Jiang}}, \bibinfo
  {author} {\bibfnamefont {C.}~\bibnamefont {Ambrosh-Draxl}}, \ and\ \bibinfo
  {author} {\bibfnamefont {M.}~\bibnamefont {Scheffler}},\ }\href@noop {}
  {\bibfield  {journal} {\bibinfo  {journal} {New J. Phys.}\ }\textbf {\bibinfo
  {volume} {14}},\ \bibinfo {pages} {023006} (\bibinfo {year}
  {2012})}\BibitemShut {NoStop}%
\bibitem [{\citenamefont {Jiang}\ \emph {et~al.}(2013)\citenamefont {Jiang},
  \citenamefont {Gomez-Abal}, \citenamefont {Li}, \citenamefont
  {Meisenbichler}, \citenamefont {Ambrosh-Draxl},\ and\ \citenamefont
  {Scheffler}}]{jiang2013}%
  \BibitemOpen
  \bibfield  {author} {\bibinfo {author} {\bibfnamefont {H.}~\bibnamefont
  {Jiang}}, \bibinfo {author} {\bibfnamefont {R.~I.}\ \bibnamefont
  {Gomez-Abal}}, \bibinfo {author} {\bibfnamefont {X.~Z.}\ \bibnamefont {Li}},
  \bibinfo {author} {\bibfnamefont {C.}~\bibnamefont {Meisenbichler}}, \bibinfo
  {author} {\bibfnamefont {C.}~\bibnamefont {Ambrosh-Draxl}}, \ and\ \bibinfo
  {author} {\bibfnamefont {M.}~\bibnamefont {Scheffler}},\ }\href@noop {}
  {\bibfield  {journal} {\bibinfo  {journal} {Comput. Phys. Commun.}\ }\textbf
  {\bibinfo {volume} {184}},\ \bibinfo {pages} {348} (\bibinfo {year}
  {2013})}\BibitemShut {NoStop}%
\bibitem [{\citenamefont {Khan}\ \emph {et~al.}(2013)\citenamefont {Khan},
  \citenamefont {Cann}, \citenamefont {Martineau}, \citenamefont {Samartseva},
  \citenamefont {Freeth}, \citenamefont {Sibley}, \citenamefont {Hartland},
  \citenamefont {Newton}, \citenamefont {Dhillon},\ and\ \citenamefont
  {Twitchen}}]{khan2013colour}%
  \BibitemOpen
  \bibfield  {author} {\bibinfo {author} {\bibfnamefont {R.~U.~A.}\
  \bibnamefont {Khan}}, \bibinfo {author} {\bibfnamefont {B.~L.}\ \bibnamefont
  {Cann}}, \bibinfo {author} {\bibfnamefont {P.~M.}\ \bibnamefont {Martineau}},
  \bibinfo {author} {\bibfnamefont {J.}~\bibnamefont {Samartseva}}, \bibinfo
  {author} {\bibfnamefont {J.~J.~P.}\ \bibnamefont {Freeth}}, \bibinfo {author}
  {\bibfnamefont {S.~J.}\ \bibnamefont {Sibley}}, \bibinfo {author}
  {\bibfnamefont {C.~B.}\ \bibnamefont {Hartland}}, \bibinfo {author}
  {\bibfnamefont {M.~E.}\ \bibnamefont {Newton}}, \bibinfo {author}
  {\bibfnamefont {H.~K.}\ \bibnamefont {Dhillon}}, \ and\ \bibinfo {author}
  {\bibfnamefont {D.~J.}\ \bibnamefont {Twitchen}},\ }\href {\doibase
  10.1088/0953-8984/25/27/275801} {\bibfield  {journal} {\bibinfo  {journal}
  {J. Phys.: Condens. Matter}\ }\textbf {\bibinfo {volume} {25}},\ \bibinfo
  {pages} {275801} (\bibinfo {year} {2013})}\BibitemShut {NoStop}%
\end{thebibliography}
\end{document}